\documentstyle[aps,epsfig]{revtex}
\begin{document}

\newcommand{\ob}{$\Omega_b$}
\newcommand{\obh}{$\Omega_bh^2$}
\newcommand{\deu}{$D$}
\newcommand{\tro}{$^3$He}
\newcommand{\qua}{$^4$He}
\newcommand{\six}{$^{6}$Li}
\newcommand{\sep}{$^{7}$Li}
\newcommand{\zaa}{{Astron. Astrophys.}}
\newcommand{\zaas}{{Astron. Astrophys. Supl.}}
\newcommand{\zapjl}{Astrophys.~J.~Lett.}
\newcommand{\zadndt}{At. Data Nucl. Data Tables}
\newcommand{\zapj}{{Astrophys. J.}}
\newcommand{\zapjs}{Astrophys.~J.~S.}
\newcommand{\znim}{{Nucl.~Inst.~and~Meth.}}
\newcommand{\znp}{{Nucl.~Phys.}}
\newcommand{\zpl}{{Phys. Lett.}}
\newcommand{\zpr}{{Phys.~Rev.}}
\newcommand{\zprl}{{Phys.~Rev.~Lett.}}
\newcommand{\znas}{New~Astronomy}
\newcommand{\znat}{Nature}
\newcommand{\zepj}{European Physical Journal}
\newcommand{\zppnp}{Progr.~Part.~Nucl.~Phys.}
\newcommand{\zmnras}{Mon. Not. R. Astron. Soc.}
\newcommand {\nc} {\newcommand}
\nc {\beq} {\begin{eqnarray}}
\nc {\eol} {\nonumber \\}
\nc {\eeq} {\end{eqnarray}}
\nc {\ve} [1] {\mbox{\boldmath $#1$}}
\nc {\vS} {\mbox{$\ve{S}$}}
\nc {\cA} {\mbox{$\cal{A}$}}
\nc{\dem} {\mbox{$\frac{1}{2}$}}
\newcommand{\henp}{\mbox{$^3$He(n,p)$^3$H}}
\newcommand{\hedp}{\mbox{$^3$He(d,p)$^4$He}}
\newcommand{\tdn}{\mbox{$^3$H(d,n)$^4$He}}
\newcommand{\hpn}{\mbox{$^3$H(p,n)$^3$He}}
\newcommand{\ddn}{\mbox{$^2$H(d,n)$^3$He}}
\newcommand{\ddp}{\mbox{$^2$H(d,p)$^3$H}}
\newcommand{\dpg}{\mbox{$^2$H(p,$\gamma$)$^3$He}}
\newcommand{\png}{\mbox{p(n,$\gamma$)$^2$H}}
\newcommand{\benp}{\mbox{$^7$Be(n,p)$^7$Li}}
\newcommand{\lipn}{\mbox{$^7$Li(p,n)$^7$Be}}
\newcommand{\lipa}{\mbox{$^7$Li(p,$\alpha)\alpha$}}
\newcommand{\tag}{\mbox{$^3$H($\alpha,\gamma)^7$Li}}
\newcommand{\heag}{\mbox{$^3$He($\alpha,\gamma)^7$Be}}
\newcommand{\sige}{\mbox{$\left( \sigma(E)\sqrt(E)\right) _0$}}

\title{Compilation and $R$-matrix analysis of Big Bang nuclear 
reaction rates}

\author{Pierre Descouvemont\thanks {Directeur de Recherches FNRS} and 
Abderrahim Adahchour\thanks {Permanent address: LPHEA, FSSM, Universit\'e Caddi Ayyad, Marrakech, Morocco
}}
\address{Physique Nucl\'{e}aire Th\'eorique et Physique Math\'ematique, CP229,\\
Universit\'{e} Libre de Bruxelles, B-1050 Brussels, Belgium}
\author{Carmen Angulo}
\address{Centre de Recherches du Cyclotron, Universit\'e catholique de Louvain,\\ 
Chemin du cyclotron 2, B-1348 Louvain-la-Neuve, Belgium}
\author{Alain Coc}
\address{Centre de Spectrom\'etrie Nucl\'eaire et de Spectrom\'etrie de
Masse, \\
CNRS/IN2P3/UPS, B\^at. 104, F-91405 Orsay Campus, France}
\author{Elisabeth Vangioni-Flam}
\address{Institut d'Astrophysique de Paris, CNRS,\\ 98$^{bis}$ Bd. Arago, 75014 Paris, France}
\maketitle
\begin{abstract}
We use the $R$-matrix theory to fit low-energy data on nuclear reactions involved in Big Bang nucleosynthesis.
A special attention is paid to the rate uncertainties which are evaluated on statistical
grounds. We provide $S$ factors and reaction rates in tabular and graphical formats.
\end{abstract}

\section{INTRODUCTION}

For a long time, Standard Big-Bang Nucleosynthesis (SBBN) was the only method
to evaluate the baryonic density in the Universe, by comparing observed and
calculated light-element abundances\cite{Oli00,Coc02} 
(\qua, \deu, \tro\ and \sep).
However, the study of Cosmic Microwave
Background (CMB) anisotropies has provided very recently a new tool for the
precise\cite{Oli03}
determination of the baryonic density, 
which can be compared to the results obtained from SBBN.
The compatibility of these two studies would lead to a more 
convincing evaluation of this fundamental cosmological parameter.
On the other hand, a significant difference would point either towards an
underestimate of the errors, or towards the need of new astrophysical models.
Since the precision on the determination of the baryonic density 
from the CMB has been drastically improved with the WMAP
satellite\cite{Spe03},
it is crucial to reduce the
uncertainties on the thermonuclear rates, which represent the main input in Standard
Big-Bang Nucleosynthesis.

Compilations of thermonuclear reaction rates for astrophysics, containing the
main reactions of SBBN, have been initiated by W.~Fowler and his collaborators.
The last version\cite{CF88} of this compilation (hereafter referred to as CF88)
concerning isotopes up to
silicon was published in 1988 but it is now partially superseded by the 
NACRE compilation\cite{NACRE}.
One of the main innovative features of NACRE with respect to former
compilations is that it provides realistic estimates of
lower and upper bounds of the rates. Using these
bounds, uncertainties on SBBN yields have been
calculated\cite{Van00a,Cyb01,Coc02}.
However, Refs. \cite{CF88} and \cite{NACRE} are broad compilations not precisely aimed at SBBN.
Compilations concerning specifically SBBN reaction rates have been performed 
by Smith, Kawano, and Malaney\cite{Smi93} (hereafter SKM) and Nollett and
Burles\cite{NB00} (hereafter NB).
They both address the main reactions of SBBN and
calculate the corresponding nuclear uncertainties.
The SKM analysis was performed using polynomial expansions for the cross 
sections, and the uncertainties on the rates were in general 
only estimated by allowing the $S$-factor limits to encompass all existing 
data, a prescription also found in some reactions covered by NACRE.
From {\em the statistical point of view}, the rate
uncertainties are better defined in the NB compilation than in SKM or
NACRE, but  the astrophysical $S$-factors of NB
are fitted by splines which have no physical justification. As the experimental cross sections
for SBBN are in general known with a fairly good accuracy, it is important not to introduce bias
due to the theoretical fit of the data.
A practical difficulty with the NB compilation is that the rates are not
provided because, by construction, they cannot be disentangled from the
Monte-Carlo calculations. 
A recent work \cite{Cyb01} also uses a subset of NACRE data limited to
the energy range of BBN (a questionable prescription)
leading to slightly different reaction rates. 

The calculation of the reaction rates is based on low-energy cross sections 
which, for charged particles, are extremely small due to the repulsive 
effect of the Coulomb barrier \cite{Cl83}. This makes measurements
in laboratories very tedious and a complementary theoretical analysis 
is in general required.
To compensate the fast energy dependence of the cross section, nuclear astrophysicists usually use the $S$-factor defined as
\beq
S(E)=\sigma(E) \, E \, \exp(2\pi\eta),
\label{eq1}
\eeq
where $E$ is the c.m.~energy, and $\eta$ the Sommerfeld parameter \cite{Cl83}.
The $S$-factor is mainly sensitive to the nuclear contribution to the cross section. For 
non-resonant reactions, its energy dependence is rather smooth.

Recent work has been focusing on primordial nucleosynthesis and on its 
sensitivity with respect to nuclear reaction rates
\cite{KR90,WSS91,Smi93,NB00,Cyb01}. In these papers, the nuclear reaction 
rates are either reconsidered by the authors
themselves \cite{Smi93}, or taken from specific works \cite{Cyb01} 
such as the Caltech \cite{CF88} or NACRE \cite{NACRE} compilations.
The goals of the present work are multiple. 
First, we analyze low-energy cross sections in the $R$-matrix 
framework \cite{LT58} which provides a more rigorous energy dependence, 
based on Coulomb functions. This approach is more complicated than those 
mentioned above, and could not be considered for broad compilations 
covering many reactions \cite{CF88,NACRE}.
However, the smaller number of reactions involved in Big Bang nucleosynthesis 
makes the application of the $R$-matrix feasible.
In addition, we do not restrict the 
data sets to the energy range of BBN, taking advantage of all data to
constrain the $S$-factor. 
A second goal of our work is a careful evaluation of the uncertainties 
associated with the cross sections and reaction rates.
This is performed here by using standard statistical techniques \cite{PDG96} and will 
be presented in more detail in Section 2.
Finally, since the completion of the NACRE compilation, several new data 
have come available (essentially data on $\henp$ \cite{BHK99} and 
$\dpg$ \cite{Lu02}) and should be included to update the reaction rates.
The reactions covered by the present analysis are:
\begin{itemize}
	\item $\dpg$
	\item $\ddn$
	\item $\ddp$
\item $\tdn$
	\item $\tag$
\item $\henp$
\item $\hedp$
\item $\heag$
\item $\lipa$
\item $\benp$
\end{itemize}
The reaction rates and $S$-factors are available at {\ttfamily http://pntpm3.ulb.ac.be/bigbang}.
We have not reconsidered the $\png$ reaction rate, for which we adopt the
analysis of Chen and Savage \cite{CS99}. The present paper deals with the
calculation of the reaction rates only.
In a separate work \cite{Van03} we analyze the consequences
of these new reaction rates on the determination of the baryonic density of
the Universe, and we will confront the results with the high precision
($\pm$4\%) value given by WMAP\cite{Spe03}. Indeed, in a previous
work \cite{Coc02} we pointed out that the compatibility between the values
obtained from CMB experiments and BBN calculations was only marginal. Thanks
to the quality of the data provided by WMAP observations, it is mandatory to
reduce drastically the nuclear uncertainties which affect the BBN 
calculations.

\section{The $R$-matrix method}
\subsection{General formalism}
Owing to the very low cross sections, one of the main problems in nuclear astrophysics is 
to extrapolate the available data down to very low energies \cite{Cl83}. Several models, such as the potential model or microscopic
approaches, are widely used for that purpose. However, they are in general not flexible enough to 
account for the data with a high accuracy. A simple way to extrapolate the data is to use a polynomial
approximation as, for example, in Ref. \cite{Smi93}. This is usually used to investigate electron screening effects, where the cross
section between bare nuclei is derived from a polynomial extrapolation of high-energy data.
This polynomial approximation, although very simple, is not based on a rigorous treatment of the
energy dependence of the cross section, and may introduce significant inaccuracies. 
As mentioned in the introduction, we use here a more rigorous approach, based on 
the $R$-matrix technique. In this method, the energy dependence of the
cross sections is obtained from Coulomb functions, as expected from the Schr\"odinger equation.
The goal of the $R$-matrix method
\cite{LT58,Th52} is to parameterize some experimentally known
quantities, such as cross sections or phase shifts, with a small
number of parameters, which are then used to extrapolate the cross
section down to astrophysical energies. 

The $R$-matrix
framework assumes that the space is divided into two regions: the internal region (with radius $a$),
where nuclear forces are important, and the external region, where the interaction between the 
nuclei is governed by the Coulomb force only. 
Although the $R$-matrix parameters do depend on the channel radius $a$, the sensitivity of the
cross section with respect to its choice is quite weak.
The physics of the internal region is parameterized by a number $N$ of poles,
which are characterized by energy $E_{\lambda}$ and reduced width
 $\tilde{\gamma}_{\lambda}$. In a multichannel problem, the $R$-matrix
at energy $E$ is defined as
\beq
R_{ij}(E) = \sum_{\lambda=1}^N \frac{\tilde{\gamma}_{\lambda i}\tilde{\gamma}_{\lambda j}}{E_{\lambda}-E},
\label{eq21}
\eeq
which must be given for each partial wave $J$. Indices $i$ and $j$  refer to the channels.
For the sake of simplicity we do not explicitly write indices $J\pi$ in the $R$ matrix and in its parameters.

Definition (\ref{eq21}) can be applied to resonant as well as to non-resonant partial waves. 
In the latter case, the non-resonant behavior is simulated by a high-energy pole, referred to as
the background contribution, which makes the $R$-matrix almost energy independent.
The pole properties
($E_\lambda$, $\tilde{\gamma}_{\lambda i}$) are known to be associated with the physical energy and width of resonances, but not strictly
equal. This is known as the difference between ``formal" parameters ($E_\lambda$, $\tilde{\gamma}_{\lambda ,i}$) and
``observed" parameters ($E^r_{\lambda}$, $\gamma_{\lambda,i}$), deduced from experiment. In a general case, involving more
than one pole, the link between those two sets is not straightforward; recent works \cite{AD00,Br02}
have established a general formulation to deal with this problem.

\subsection{Elastic scattering}
Elastic scattering does not directly present an astrophysical interest, but is the basis
for capture and transfer reactions. In single-channel calculations, the $R$ matrix is a function which is given by
\beq
R(E) = \sum_{\lambda=1}^N \frac{\tilde{\gamma}_{\lambda}^2}{E_{\lambda}-E},
\label{eq22}
\eeq
and the phase shift is given by
\beq
\delta=\delta_{\rm HS}+\delta_{\rm R}=-\arctan\frac{F(ka)}{G(ka)}+\arctan \frac{P R}{1- S R},
\label{eq23}
\eeq
where we have introduced the hard-sphere phase shift $\delta_{\rm HS}$ and the $R$-matrix
phase shift $\delta_{\rm R}$.
In eq. (\ref{eq23}), $k$ is the wave number and $F$ and $G$ are the Coulomb functions (we do not
explicitly write the angular momentum $\ell$); the penetration and shift factors $P$ and $S$ 
are given by
\beq
L  =  k a \, \frac{O^{\prime}(ka)}{O(ka)} = S + i P,
\label{eq25}
\eeq
where the outgoing Coulomb function $O$ is given by $O=G+iF$ \cite{LT58}.

The link between formal and observed parameters is discussed, for example, in 
Refs. \cite{LT58,AD00,Br02}. Here we just mention the main results.
The resonance energy $E^r_i$, or the ``observed" energy, is defined as the energy where the $R$-matrix phase shift
is $\delta_R=\pi/2$. According to (\ref{eq23}), $E_i^{\rm r}$ is therefore a solution of the equation
\beq
S(E_i^{\rm r})R(E_i^{\rm r}) = 1.
\label{eq26}
\eeq

If the pole number $N$ is larger than unity, the link between observed and calculated parameters is not analytical
and requires numerical calculations \cite{AD00}.
We illustrate here the simple but frequent situation for $N=1$, where
\beq
&&E_1^{\rm r}=E_1-\tilde{\gamma}_1^2 S(E_1^{\rm r}), \nonumber \\
&&\gamma_1^2=\tilde{\gamma}_1^2/\left(1+\tilde{\gamma}_1^2\, S'(E_1^{\rm r}) \right),
\label{eq210}
\eeq
with $S'(E)=dS/dE$. These formulas provide a simple link between calculated and observed values.
In eq.(\ref{eq210}), $\gamma_1^2$ is the observed reduced width, defined from the experimental
width $\Gamma_1$ by the well known relationship
\beq
\Gamma_1=2 \gamma_1^2 P(E_1^{\rm r}).
\label{eq210b}
\eeq
Equations (\ref{eq210}) allow to determine the $R$ matrix parameters from the experimental data.

\subsection{Transfer reactions}
Let us consider two colliding nuclei with masses ($A_1$, $A_2$), charges ($Z_1e$, $Z_2e$) and spins ($I_1$, $I_2$).
The transfer cross section $\sigma_t(E)$ from the initial state to a final state is defined as
\beq
\sigma_t(E)=\frac{\pi}{k^2}(1+\delta_{12})\sum_{J\pi}\frac{2J+1}{(2I_1+1)(2I_2+1)}
\sum_{\ell \ell' I I'} |U^{J\pi}_{\ell I,\ell'I'}(E)|^2,
\label{eq211}
\eeq
where $\delta_{12}$ is 1 or 0, for symmetric and non-symmetric systems, respectively.
The collision matrix $\ve{U}^{J\pi}(E)$
contains the information about the transfer process. Quantum numbers ($\ell I$) and ($\ell' I'$)
 refer to the entrance and exit channels, respectively. In general, for given total angular 
momentum $J$ and parity $\pi$, several $I$ values (arising from the coupling of $I_1$ and $I_2$) 
and $\ell$ values are allowed.
To simplify the presentation, we assume here that a single set of ($\ell I$) and ($\ell' I'$) values is involved in 
(\ref{eq211}). This is justified at low energies where the lowest angular momentum is strongly dominant.

As shown in ref.\cite{AD98}, the collision matrix $\ve{U}$ is deduced from the $R$-matrix by
\beq
U_{11} & = & \frac{I_1}{O_1}\, \frac{1-R_{11}L_1^{\star}-R_{22}L_2}{1-R_{11}L_1-R_{22}L_2}, \\ 
&&  \nonumber \\                                                                                          
U_{22} & = & \frac{I_2}{O_2}\, \frac{1-R_{11}L_1-R_{22}L_2^{\star}}{1-R_{11}L_1-R_{22}L_2},\nonumber \\  
&&\nonumber \\                                                                                           
U_{12} & = & U_{21} = \frac{2ia \sqrt{k_1k_2}\sqrt{R_{11}R_{22}}}{O_1\,O_2\,(1-R_{11}L_1-R_{22}L_2)},\nonumber \\ 
\label{eq212}
\eeq
where we have introduced the incoming Coulomb functions $I_1$ and $I_2$, defined by $I_i=O^{\star}_i$. In these equations, the Coulomb functions are evaluated at the channel radius $a$. 
When a single pole is present, eq. (\ref{eq210}), defined for single-channel systems,
is extended to
\beq
E_1^{\rm r}&=&E_1-\tilde{\gamma}_{1,1}^2 S_1(E^{\rm r}_1) -\tilde{\gamma}_{1,2}^2 S_2(E^{\rm r}_1),\nonumber \\
\gamma_{1,i}^2&=&\tilde{\gamma}_{1,i}^2/\left(1+\tilde{\gamma}_{1,1}^2\, S'_1(E^{\rm r}_1)+\tilde{\gamma}_{1,2}^2\, S'_2(E^{\rm r}_1) \right).
\label{eq215}
\eeq

If no resonance is present in the energy range of interest, the $R$-matrix (\ref{eq21}) involves high-energy 
poles only. In that case it can be parameterized by a constant value
\beq
R_{ij}(E)=R^0_{ij},
\label{eq213}
\eeq
with the constraint 
\beq
(R^0_{12})^2=R^0_{11}\, R^0_{22}
\label{eq214}
\eeq
if a single pole is involved.

\subsection{Radiative-capture cross sections}
The determination of capture cross sections requires the calculation of matrix elements of 
the multipole operators ${\cal M}^{\sigma}_{\lambda}$.
According to the $R$-matrix method, such a matrix element between two wave functions $\Psi_i$ and $\Psi_f$
is written as
\beq
<\Psi_f|| {\cal M}^{\sigma}_{\lambda} || \Psi_i>=<\Psi_f|| {\cal M}^{\sigma}_{\lambda} || \Psi_i>_{int}
+<\Psi_f|| {\cal M}^{\sigma}_{\lambda} || \Psi_i>_{ext}={\cal M}_{int}+{\cal M}_{ext},
\label{eq216}
\eeq
where ${\cal M}_{int}$ and ${\cal M}_{ext}$ represent the internal and external contributions,
respectively. Wave function $\Psi_f$ corresponds to the final (bound) state whereas $\Psi_i$ describes the
initial scattering state at energy $E$. In the internal region their effect is simulated by
the pole properties \cite{LT58}. At large distances, their asymptotic behaviors are given by
\beq
\Psi_f \rightarrow C_f W_{-\eta_f,\ell_f+1/2}(2k_f\rho) \nonumber \\ 
\Psi_i \rightarrow I_{\ell_i}(k\rho)-U^{\ell_i}O_{\ell_i}(k\rho), 
\label{eq217}
\eeq
where $C_f$ and $k_f$ are the Asymptotic Normalization Constant (ANC) and the wave 
number of the final wave function, respectively; $W$ is the
Whittaker function, and $U^{\ell_i}$ is the collision matrix of the initial state.

The capture cross section is then defined as
\beq
\sigma_c=\frac{\pi}{k^2}(1+\delta_{12})\sum_{J_i,\pi_i}\frac{2J_i+1}{(2I_1+1)(2I_2+1)}
|U^{\gamma}(J_i\pi_i \rightarrow J_f \pi_f)|^2,
\label{eq218}
\eeq
which extends the transfer cross section (\ref{eq211}) to reactions involving photons \cite{LT58}.
The ``equivalent" collision matrix is divided in two parts
\beq
U^{\gamma}=U^{\gamma}_{int}+U^{\gamma}_{ext}.
\label{eq219}
\eeq
The internal part $U^{\gamma}_{int}$ is written as
\beq
U^{\gamma}_{int}=i^{\ell}\exp(i\delta_{\rm HS})  \frac{1}{1-LR} 
\sum_{i} \frac{ \sqrt{\tilde{\Gamma}_i \tilde{\Gamma}_{\gamma,i}}}{E_i-E}
\label{eq220}
\eeq
where we have defined a further pole parameter, the gamma width of pole $i$, as
\beq
\tilde{\Gamma}_{\gamma,i}=\frac{8\pi(\lambda+1)k_{\gamma}^{2\lambda+1}}{\lambda (2\lambda+1)!!^2}  
\,\frac{2J_f+1}{2J_i+1} |<\Psi_f|| {\cal M}^{\sigma}_{\lambda} || \varphi_i>_{int}|^2.
\label{eq221}
\eeq

For electric multipoles $U^{\gamma}_{ext}$ is given by
\beq
U^{\gamma}_{ext}&=&i^{\ell +1}e
C_f F_E \sqrt{\frac{2J_f+1}{2J_i+1} \, \frac{8 \pi(\lambda+1)k_{\gamma}^{2\lambda+1}}{\hbar v \lambda (2\lambda+1)!!^2}}\nonumber \\
& & \times \int_a^{\infty} W_{-\eta_f,\ell_f+1/2}(2k_f\rho) 
\left[ I_{\ell_i}(k\rho)-U^{\ell_i}O_{\ell_i}(k\rho) \right] \rho^{\lambda} \, d\rho,
\label{eq222}
\eeq
where the geometrical factor $F_E$ reads
\beq
F_E&=&  \left[ Z_1 \left( \frac{A_2}{A} \right)^{\lambda}+Z_2 \left( \frac{-A_1}{A} \right) ^{\lambda}
\right] \left[ \frac{(2\lambda+1)(2J_i+1)(2\ell_i+1)}{4\pi} \right] ^{\dem} <\ell_i \, 0 \, \lambda \, 0|\ell_f \, 0>
\left \{ \begin{array}{ccc} J_i & J_f & \lambda \\ \ell_f & \ell_i & I \end{array} \right \},
\label{eq223}
\eeq
where $I$ is the channel spin, coming from the coupling of $I_1$ and $I_2$.
This presentation is general. In the present work, none of the capture reactions involves a
resonance at low energy. Consequently, the internal contribution (\ref{eq220}) is determined
with a single pole at energy $E_0$, which simulates the background.

\section{Treatment of uncertainties}
Improvements of the current work on Big Bang nucleosynthesis essentially concerns a more 
precise evaluation of uncertainties on the reaction 
rates. Here, we address this problem by using standard statistical methods \cite{PDG96}.
This represents a significant improvement with respect to NACRE \cite{NACRE}, where
uncertainties are evaluated with a simple prescription, necessary for a simultaneous
analysis of many reactions, but which does not correspond to a rigorous statistical
treatment. The NACRE error bars should not be interpreted as $1\sigma$ errors.
The $R$-matrix approach depends on a number of parameters, some of them being fitted, whereas others are constrained 
by well determined data, such as energies or widths of resonances. Let us denote by $\nu$ the number of free parameters $p_i$. The choice of the free parameters is guided by the physics of the problem.
The reduced $\chi^2$ value is defined as
\beq
\chi^2(p_i)=\frac{1}{N-\nu} \sum_{k=1}^{N}\left(
\frac{\sigma_k^{exp}-\sigma_k^{th}(p_i)}{\Delta\sigma_k^{exp}} \right) ^2
\label{eq301}
\eeq
where $N$ is the number of experimental data, $\sigma^{exp}_{k}$ is the experimental cross section (with uncertainty
$\Delta\sigma^{exp}_{k}$) and $\sigma^{th}_{k}(p_i)$ the $R$-matrix cross section at the corresponding energy.
As usual, the adopted parameter set is obtained from the minimal $\chi^2$ value. 
Notice that this definition assumes that the data sets are independent of each other. A more
general definition, involving the covariance matrix can be found in Ref.\cite{PDG96}.
However, with the currently available data, Eq. (\ref{eq301}) should be used, which could
slightly affect the quoted uncertainties.
The uncertainties on the parameters are
evaluated as explained in Ref.\cite{PDG96}. The range of acceptable $p_i$ values is such that
\beq
\label{eq302}
\chi^2(p_i) \leq \chi^2(p_i^{min}) + \Delta \chi^2,
\eeq
where $p^{min}_{i}$ is the optimal parameter set. In this equation, $\Delta \chi^2$ is obtained from
\beq
P(\nu/2, \Delta \chi^2/2)=1-p,
\label{eq303}
\eeq
where $P(a,x)$ is the Incomplete Gamma function, and $p$ is the confidence limit ($p=0.683$ for the $1\sigma$
confidence level). We refer to Refs. \cite{PDG96,NUM} for details. Equation (\ref{eq302}) defines a region
of $R$-matrix parameters acceptable for the cross-section fits. This range is scanned for all parameters, and the limits on the
cross sections are then estimated at each energy. 

The optimal parameters are complemented by the covariance matrix (Table II). 
The covariance matrix $\ve{C}$ between parameters $p_i$ and $p_j$ is defined \cite{NUM} from
\beq
C_{ij}=[\ve{\alpha}^{-1}]_{ij} \nonumber \\
\alpha_{ij}=\frac{1}{2}\frac{\partial^2\chi^2}{\partial p_i\partial p_j}, 
\label{cov}
\eeq
where the derivatives are calculated at the minimum. Uncertainties on parameters $p_i$ are
determined from
\beq
\Delta p_i=\sqrt{C_{ii}},
\eeq
but, in general the off-diagonal elements are large, and individual errors cannot be quoted without
the covariance matrix.

In many references, the statistical and systematic errors are not available separately. As we
want to use an homogeneous treatment, we have combined then in the fitting procedure, when
available. An advantage of the physical energy dependence provided by the $R$-matrix formalism is
that it gives a further constraint on the fit. Consequently the recommended uncertainties can
be lower than the systematic uncertainty. This would not be true with polynomial approximations
for example, where the resulting fits must be scaled by the systematic error.

As is well known, several reactions involved in nuclear astrophysics present different data sets which are not compatible
with each other. An example is the $\heag$ reaction where data with different normalizations are available.
In this case we have used a procedure adapted from the recommendations of Audi and Wapstra \cite{AW95} and of the Particle
Data Group \cite{PDG96}.
\\

(i) \underline{Case 1 : $\chi^2 >1$}

If no systematic difference exists between the normalizations, and if $\chi^2$ is significantly larger than 1, this means that the error bars 
have been underevaluated in the original work. This would give recommended $S$ factors with too low uncertainties. According to refs \cite{AW95,PDG96},
the errors bars of the data with the individual $\chi_k^2$ defined by
\beq
\chi^2_k=\left(
\frac{\sigma_k^{exp}-\sigma_k^{th}(p_i)}{\Delta\sigma_k^{exp}} \right) ^2 >1
\label{eq304}
\eeq
have been multiplied here by $\sqrt{\chi^2_k}$. 
In this way, the global $\chi^2$ value (\ref{eq301}) is equal to 1, and the usual method can be
used to evaluate the uncertainties on the $S$ factor.
\\ \\

(ii) \underline{Case 2 : different normalizations}

In some reactions, the differences between data sets obviously arise from different normalizations. The standard $\chi^2$
method is meaningless in this case since: $(i)$ the $\chi^2$ value is most likely larger than 1; $(ii$) the weight of data sets with many
data is overestimated, compared to data sets with less data. In those circumstances, we have performed individual
fits of each data set separately. The procedure is detailed below.
\begin{itemize}
\item Step 1\\
Each data set is fitted individually (Fig. 1, panel (a)). Then extrapolation of all 
sets provides the cross sections 
at any energy (Fig. 1, panel (b)).
\item Step 2\\
At a given energy $E_k$, an averaged cross section is determined as
\beq
\sigma^{eff}(E_k)=\frac{\sum_{i=1}^{N_{exp}}\sigma_i(E_k)/\Delta \sigma_i^2}
{\sum_{i=1}^{N_{exp}}1/\Delta \sigma_i^2},
\label{eq305}
\eeq
where $N_{exp}$ is the number of data sets and $\Delta\sigma_i$ the error bar (for extrapolated data the error bar is taken 
as the largest value). This step is shown in Fig. 1 (panel (c)). Along with Eq. (\ref{eq305}) 
an effective error bar is determined as 
\beq
\Delta \sigma^{eff}=1/\sqrt{\sum_{i=1}^{N_{exp}}1/\Delta \sigma_i^2}
\label{eq306}
\eeq
\item Step 3\\
At energy $E_k$, a partial $\chi^2$ is determined, and the error bar (\ref{eq306}) is renormalized if $\chi^2>1$.
This method provides a reasonable way to deal with data sets presenting different normalizations. 
It has been used for the $\ddn$ and $\ddp$ reactions. 
\end{itemize}
\begin{figure}[h]
\centerline{\epsfig{file=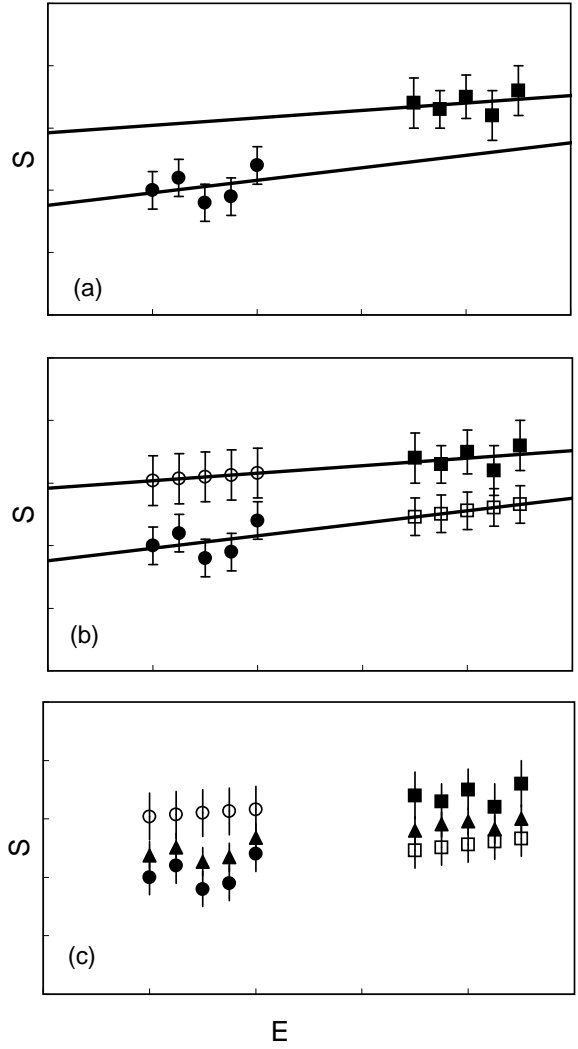,width=9cm,clip=}}
\caption{Illustration of case 2 (see text). Full circles and squares represent the original
experimental data (units are arbitrary); open symbols are extrapolated values, and triangles represent the
``effective" cross sections (\ref{eq305}).
\label{fig2}}
\end{figure}

\section{Calculation of Reaction rates}
\subsection{Definition}
The reaction rate $N_{\rm A}\langle\sigma v\rangle$ is defined as \cite{FO67}
\begin{eqnarray}
N_{\rm A}\langle\sigma v\rangle & = & 
N_{\rm A}\ \frac{(8/ \pi)^{1/2}}{\mu^{1/2}(k_{\rm B}T)^{3/2}}
\int^{\infty}_{0} \sigma E \exp(-E/k_{\rm B}T) \ dE,
\label{equ1}
\end{eqnarray}
where $N_{\rm A}$ is the Avogadro number, $\mu$ the reduced mass of
the system, $k_{\rm B}$ 
the Boltzmann constant, $T$ the temperature, $\sigma$ the cross section,
$v$ the relative velocity, and $E$ the energy in the centre-of-mass system.  
When $N_{\rm A} \langle \sigma v \rangle$ is expressed in 
cm$^3$ mol$^{-1}$ s$^{-1}$, the energies $E$ and $k_{\rm B}T$ 
in MeV, and the cross section $\sigma$ in barn, Eq.~(\ref{equ1})
leads to
\begin{eqnarray}
N_{\rm A}\langle\sigma v\rangle & = & 
3.7313\times10^{10}\ \mu_0^{-1/2} \, T_9^{-3/2}
\int^{\infty}_0 \sigma \, E \, \exp(-11.604 E/T_9) \, dE,
\label{equ2}
\end{eqnarray}
where $\mu_0$ is the reduced mass in amu, and $T_9$ is the temperature in units of $10^9$ K. 
The calculation of the rates is performed here between $T_9 = 0.001$ and 10, and is compared
with previous compilations \cite{CF88,NACRE}. 

Let us first discuss charged-particle reactions. Except near narrow resonances, the $S$-factor is a 
smooth function of energy, which is convenient for extrapolating
measured cross sections down to astrophysical energies.
When $S(E)$ is assumed to be a constant, the integrand in Eq.~(\ref{equ1}) 
is peaked at the ``most effective energy" (the Gamow energy \cite{FO67}),

\begin{eqnarray}
E_0  =  \left(\frac{\mu}{2}\right)^{1/3}
\left( \frac{\pi e^2 Z_1Z_2 k_{\rm B}T}{\hbar}\right)^{2/3} 
 = 0.1220\ (Z_1^2Z_2^2 \mu_0)^{1/3} \, T_9^{2/3}\  {\rm MeV}
\label{equ5}
\end{eqnarray}
and can be approximated by a Gaussian function centered at $E_0$, with 
full width at $1/e$ of the maximum given by

\begin{eqnarray}
\Delta E_0  =  4\, (E_0\ k_{\rm B}T/3)^{1/2}  
 = 0.2368 \, (Z_1^2 Z_2^2 \mu_0)^{1/6}\, T_9^{5/6} \ {\rm MeV}. 
\label{equ6}
\end{eqnarray}

With these approximations, the integral in Eq.~(\ref{equ1}) can be 
calculated analytically \cite{FO67}. However, in the present compilation we
do not rely on such approximations and perform numerically the integration
of Eq.~(\ref{equ1}). A good accuracy is reached by limiting the 
numerical integration for a given temperature to the energy domain 
($E_0-n\Delta E_0,E_0+n\Delta E_0$), with typically $n=2$ or 3. The accuracy is such
that at least 4 digits on the rate are significant. For neutron-induced reactions, Eq.~(\ref{equ1}) is integrated numerically from
$E=0$ to $E=nk_BT$, where $n$ is typically 10.

Table V presents the rate in a numerical format. 
To interpolate we recommend to following procedure. As is it well known \cite{CF88}, the non-resonant
reaction rate can be parametrized as 
\beq
N_{\rm A}\langle\sigma v\rangle=\exp(-C_0/T^{1/3})f_0(T)/T^{2/3},
\eeq
where $C_0$ only depends on masses and charges of the system, and is defined by
\beq
C_0=3\left[ \pi \frac{e^2}{\hbar c} Z_1Z_2(\frac{\mu c^2}{2k_B})^{1/2} \right] ^{\frac{2}{3}},
\eeq
and where $f_0(T)$ is a smooth function of $T$. Interpolating $f_0(T)$ or $\log \, f_0(T)$
with a spline method provides the rates with a good accuracy (typically better than 0.1\%).

\subsection{Screening effects}
In stellar plasmas, atoms are usually completely ionized, and nuclear reactions involve bare nuclei.
The situation is different in laboratories since target nuclei are partially- or un-ionized.
Consequently the role of the electron cloud cannot be neglected at low energies. Let us notice
that screening effects, with a different origin, may also occur in stars, but this issue is far
beyond our topic.

The screening effect is usually evaluated through the screening potential $U_e$.
The screening factor \cite{ALR87} is defined as
\beq
f(E)=\frac{\sigma_{exp}(E)}{\sigma_{th}(E)}=\frac{\sigma_{th}(E+U_e)}{\sigma_{th}(E)}
\approx \exp(\pi\eta\frac{U_e}{E}),
\label{eq250}
\eeq
where $\sigma_{exp}(E)$ is the experimental cross section, affected by screening effects, and 
${\sigma}_{th}(E)$ the theoretical cross section involving bare nuclei. Here the $R$-matrix fit
has been applied at energies unaffected by screening effects, and a screening potential has
been deduced. For an extended $R$-matrix analysis of electron-screening effects, see for
example ref. \cite{Ba02}.

\subsection{Physical constants}
In the analysis of the cross sections and in the calculation of the reaction rates, we have used 
the atomic masses as recommended by Audi {\sl et al.} \cite{ABB97}. The following values of the physical constants are used:
\beq
c & = & 299792458 \ {\rm m \, s}^{-1} \nonumber \\
1\ amu & = & 931.494\ {\rm MeV}/c^2 \nonumber \\
1\ {\rm eV} & = & 1.60218\times 10^{-19} \ {\rm J} \nonumber  \\
k_{\rm B}T & = & 0.08617\,T_9 = T_9/11.605\ {\rm MeV} \nonumber \\
N_{\rm A} & = & 6.0221\times 10^{23}\ {\rm mol}^{-1} \nonumber \\
\alpha & = & e^2/\hbar c  =  1/137.036 \nonumber \\
\hbar c & = & 197.327 \ {\rm MeV.fm} \nonumber
\eeq

\section{Cross sections and reaction rates}
\noindent
{\em $\dpg$}\\
The data of ref. \cite{Sc95} are superseded by ref. \cite{Sc97} and are therefore not
included.  Very recent data \cite{Lu02} allow a more precise extrapolation
down to low energies. Below 0.01 MeV, the $S$ factor is nearly constant which is typical
of $s$-wave capture, proceeding by an M1 transition. At zero energy, our partial $S$ factors 
$0.089 \pm 0.004$ eV.b (E1) and $0.134 \pm 0.006$ eV.b (M1) are consistent with the
values recommended by Schmid {\sl et al.} \cite{Sc97} ($0.073$ eV.b  and $0.109$ eV.b,
respectively) from polarized-data measurements. Our results are slightly higher than NACRE,
which uses a polynomial fit for the $S$ factor.\\

\noindent
{\em $\ddn$ and $\ddp$}\\
Two non-resonant partial waves are included in the fit. The fits have been performed
individually (data of refs. \cite{KR87a,BR90,GR95}), with each of them being complemented
by the high-energy data of ref. \cite{SC72}. The recommended $S$ factors have been deduced
as explained in Sec. III. The individual fits are given in figs \ref{ddn} and \ref{ddp} as dotted lines. As shown in Ref.\cite{AD98} it is not possible to optimize the fits of both reactions with
the same parameter set. Consequently the $R$-matrix parameters are somewhat different.
The reaction rates are close to the results of NACRE, but the uncertainties have been
reduced.\\

\noindent
{\em $\tdn$}\\
In addition to the well known low-energy $3/2^+$ resonance ($\ell_i=0$), non-resonant
contributions from the $1/2^+\ (\ell_i=0)$ and $1/2^-,3/2^-\ (\ell_i=1)$ partial waves
have been included. The present $R$-matrix fit is very close to the fit of Hale \cite{Hale},
and yields a fairly low uncertainty on the reaction rate. We find a reaction rate similar
to NACRE, except at high temperatures, where NACRE uses very conservative lower and upper bounds.\\

\noindent
{\em $\tag$}\\
The data of refs. \cite {GR61,SC87a} have not been included as they are obviously
inconsistent with the other data sets. The $s-$ and $d$-wave contributions are taken into account. 
To reduce the number of free parameters, we have adopted, for $\tag$ and $\heag$, the same
ANC values for the ground and first excited states. This seems reasonable as both states arise
from the same isospin doublet.
The statistical method adopted here provides error bars significantly lower
than in NACRE, where a very conservative technique was used. At high energies, the
reaction rates are slightly different; in spite of the lack of data above 1.2 MeV,
the $R$-matrix approach is expected to be more reliable than the polynomial extrapolation
used in NACRE.\\

\noindent
{\em $\henp$}\\
In the low-energy region, the main partial waves correspond to $\ell= 0$ and 1. According to the
$^4$He energy spectrum, 
the $0^+_2\ (E_x=20.21$ MeV), $0^-\ (E_x=21.01$ MeV) and $2^-\ (E_x=21.84$ and 23.33 MeV) states
are expected to determine
the cross section. They correspond to ($\ell I) = (0,0), (1,1)$ and (1,1), respectively. The role of the two broad 
$2^-$ resonances has been simulated by a single pole in the $R$-matrix expansion. An $\ell=0$ 
non-resonant partial wave, corresponding to $J=1^+$ has been also taken into account. 
The data
of Brune {\sl et al.} \cite{BHK99} suggest a new $0^-$ resonance at 0.43 MeV which indeed must
be included to optimize the fit.
More detail can be found in Ref. \cite{AD03}.\\

\noindent
{\em $\hedp$}\\
The dominance of the $\ell=0$ contribution at low energy is confirmed by the isotropic
angular distributions \cite{KR87}, but a ${\ell_i=1}$ component has been included to improve
the quality of the fit. The Coulomb dependence involved in
the $R$ matrix approach leads to differences up to 10\% with the polynomial expansion used by Krauss
{\sl et al.} \cite{KR87}.
This explains the differences with previous compilations \cite{CF88,NACRE}. Our rate
is in good agreement with the fit of Hale \cite{Hale}.\\
The low-energy data of Refs. \cite{Co00} and \cite{Al01} are affected 
by electron-screening effects. The former are obtained through the d($^3$He,p)$^4$He reaction,
and are complemented by a subset of the latter data \cite{Al01}. The screening potentials are 
found as $U_e=146 \pm 5$ eV for the d($^3$He,p)$^4$He reaction,
and $U_e=201 \pm 10$ eV for the $\hedp$ reaction. These values are somewhat different from those
derived in ref.\cite{Al01} ($109 \pm 9$ eV and $219 \pm 7$ eV) where a polynomial approximation
is used to determine the bare-nucleus cross sections. According to ref. \cite{Al01}, we do not include
the data of refs. 
\cite{En88,Pr94} as their analysis was biased by stopping-power problems.\\

\noindent
{\em $\heag$}\\
A purely external capture has been assumed, with $\ell_i=0$ and $\ell_f=2$ contributions.
The data of ref. \cite{Na69} are clearly affected by normalization problems, and have not been taken into account. According to Ref. \cite{Hi88}, the data of Kr\"awinkel {\sl et al.} \cite{KR82} have
been renormalized by 1.4.
For this reaction, most data sets allow an
extrapolation down to zero energy. Accordingly, an $S(0)$ value, with the associated uncertainty,
has been determined for each reaction, and an averaged $S(0)$ has been obtained. Since the capture cross section is assumed
to be external, the $S$ factor only depends on the normalization factor. The
normalization has been deduced from the adopted $S(0)$. The present $S(0)$ value 
($S(0)=0.51 \pm 0.04$ keV.b) overlaps with the value recommended by Adelberger {\sl et al.}
\cite{AD98b} ($S(0)=0.53 \pm 0.05$ keV.b) and by NACRE \cite{NACRE} ($S(0)=0.54 \pm 0.09$ keV.b). 
\\

\noindent
{\em $\lipa$}\\
The $S$ factor is mainly determined by $\ell_i=1, J=0^+,2^+$ contributions. Owing to parity
conservation and to the symmetry of the final state, $\ell=0$ partial waves in the
entrance channel are forbidden. The $^8$Be spectrum presents two $2^+$ states below
the $^7$Li+p threshold. These states have been accounted for by a single state at $E=-0.48$
MeV. For the $2^+$ resonance at $E=2.60$ MeV, we neglect the interference with the
subthreshold state; the energy and widths have been taken from literature, without
any fitting procedure.\\
At very low energies, data affected by electron screening ($E < 40$ keV) have not been considered 
in the fitting procedure. An analysis of the screening potential provides $U_e=100 \pm 25$ eV. This
value is much lower than the value deduced by Engstler {\sl et al.} ($U_e=300\pm 280$ eV for an
atomic target, $U_e=300\pm 160$ eV for a molecular target) who use a third-order polynomial
to determine the bare-nucleus cross section. This procedure is quite questionable here since
the low-energy $S$-factor depends on a subthreshold state whose effect is negligible beyond 100 keV.
A recent
experiment by Lattuada {\sl et al.} \cite{LA01} uses the Trojan Horse Method which does not
depend on electron screening, and provides  $S(0)=55\pm 3$ keV.b by a polynomial extrapolation. The present analysis provides a significantly higher $S$ factor at
low energy ($S(0)=67\pm 4$ keV.b). This discrepancy is confirmed by a recent $R$-matrix analysis
of Barker \cite{Ba00,Ba02} who finds values similar to ours.\\

\noindent
{\em $\benp$}\\
The $2^-$ state located very near threshold determines the cross section in a wide energy range.
To reproduce the data up to 5 MeV we have included the $3^+$ resonances at $E=0.34$ MeV
and $E=2.60$ MeV. We neglect interference effects.
Our reaction rate is consistent with the SKM compilation up to
$T_9 \approx 4 $, but provides larger values above this temperature.
More detail can be found in Ref. \cite{AD03}.

\section*{Acknowledgments}
We are grateful to Jeff Schweitzer for useful comments on the manuscript, and to Carl Brune for providing us with the $\henp$ data in a numerical format. A.A. thanks the
FNRS for financial support. This text presents research results of the Belgian program P5/07 on
interuniversity attraction poles initiated by the Belgian-state
Federal Services for Scientific, Technical and Cultural Affairs.

\newpage
\section*{Explanation of tables}
\begin{description}
	\item[TABLE I] {\bf $R$-matrix parameters (observed values).}\\
	The observed values are given.
	The channel radius $a$ is taken as $a=5$ fm, except for the $\tag$ and $\heag$ reactions,
	where $a=3$ fm. Non-fitted parameters are shown in italics.\\
	{\sl Capture reactions:}\\
	$\ell_i,J_i$: orbital momentum and total spin of the initial state.\\
	$\ell_f,J_f,E_f,C_f$: orbital momentum, total spin, energy and ANC of the final state
	($E_f$ is taken from literature).\\
	$E_1^r,\Gamma_i,\Gamma_f$: $R$-matrix parameters (see Sec.II).\\ \\
	{\sl Transfer reactions:}\\
		$\ell_i,J_i$: orbital momentum and total spin of the initial state.\\
	$\ell_f$: orbital momentum of the final state. \\
	$E_1^r,\Gamma_i,\Gamma_f$: $R$-matrix parameters for resonant partial waves. \\
	$R_i,R_f$: $R$-matrix parameters for non-resonant partial waves. \\

	\item[TABLE II] {\bf Covariance matrices.}\\
	The covariance matrices $\ve{C}$ are calculated from Eq.(\ref{cov}). Units are chosen as in Table I.
	
	\item[TABLE III] {\bf Zero-energy $S$ factors (or $\sige$ for neutron-capture reactions).}
	\item[TABLE IV.a-b] {\bf $S$-factors.}\\
	Energies are chosen from zero to the experimental upper limits, with a step which provides
	an accurate interpolation.
	\item[TABLE V] {\bf Analytical fits of the reactions rates.}\\
	$T_{9}^{max}$: maximum value of $T_9$ for which the fit reproduces the numerical values of
	Tables VI with an accuracy better than 5\%.\\
	The parametrization is as follows:
	\beq
	N_{\rm A}\langle\sigma v\rangle&=&d_0 \ \frac{\exp(-C_0/T_9^{1/3})}{T_9^{2/3}}\times(1+\sum_{i=1}^3 d_i T_9^i) 
	{\rm \ for\ charged\ particles} \nonumber \\
	N_{\rm A}\langle\sigma v\rangle&=&d_0 \ (1+\sum_{i=1}^3 d_i T_9^i) {\rm \ for\ neutrons},
	\nonumber 
\eeq	
with units:\\
 $T_9: 10^9\, K$,\\
  $C_0: K^{1/3}$,\\
$N_{\rm A} \langle \sigma v \rangle, d_0:$ cm$^3$ mol$^{-1}$ s$^{-1}$,\\

	\item[TABLE VI.a-j] {\bf Reaction rates (in cm$^3$ mol$^{-1}$ s$^{-1}$).}\\
	$E_0,\Delta E_0$ (in MeV): see Eqs. (\ref{equ5}) and (\ref{equ6})\\
 NACRE: ratio of the adopted rate with respect to the NACRE rate \cite{NACRE} \\
 SKM: ratio of the adopted rate with respect to the SKM rate \cite{Smi93} \\
 Lower and upper values correspond the one-sigma uncertainties.
\end{description}

\section*{Explanation of graphs}
\begin{description}
\item[Figures I.a-j] {\bf $S$-factors}\\
The figures represent the $S$ factors for charged particles, and $\sigma(E)\sqrt{E}$
for neutron induced reactions (full curves), versus c.m. energy. If not specified, the dotted 
curves represent the lower and upper limits.

\item[Figures II.a-b] {\bf Reaction rates}\\
Reaction rates normalized to the NACRE adopted rates, or
to the SKM rates for the $\hedp$, $\henp$, and $\benp$ reactions not
available in NACRE.
Solid curves correspond to the present reaction rates, dashed curves to the
SKM rates and dotted curves to the NACRE upper and lower limits.
\end{description}

\newpage
\begin{table}[h]
\caption{$R$-matrix parameters (observed values). Energies and widths are expressed in MeV and ANC in fm$^{-1/2}$. $R$ matrices are dimensionless. The reduced $\chi^2$ values are given in brackets. }
\begin{tabular}{llllllllll}
Capture reactions & $\ell_i$ & $J_i$ & $\ell_f$ & $J_f$ & $E_f$ & $C_f$  & $E^r_1$ & $\Gamma_i$ & $\Gamma_f$ \\ 
\hline
$\dpg$ (0.55)& 1 & $3/2^-$ & 0 & $1/2^+$ & -5.49 & -1.78 & {\sl 10} & 23.4 & $4.20\times 10^{-4}$  \\
& 0 & $1/2^+$ & 0 & $1/2^+$ & -5.49 & -1.78& {\sl 10} & {\sl 23.4} & $7.28\times 10^{-5}$   \\
&&&&&&&&& \\
 $\tag$ (0.16)& 0 &  $1/2^+$  & 1 &  $3/2^-$  & -2.47& -3.49  & {\sl 10} & 4.87 & $1.71\times 10^{-2}$ \\
 & 2$^{a)}$ &  $5/2^+$  & 1 &  $3/2^-$  & -2.47 & -3.49&   &   &    \\
 & 0 &  $1/2^+$  & 1 &  $1/2^-$  & -1.99& -3.49 & {\sl 10} & {\sl 4.87} & $3.70\times 10^{-3}$   \\
 & 2$^{a)}$ &  $3/2^+$  & 1 &  $1/2^-$  & -1.99& -3.49 &  &   &     \\
\end{tabular}
$^{a)}$ External capture only.\\
$^{b)}$ Reduced width $\gamma_i^2$.
\end{table}

\newpage
\begin{table}[h]
\caption{Covariance matrices. }
\begin{tabular}{l|ccccccccc}
Capture reactions \\
\hline
 $\dpg$  &$\Gamma_i(3/2^-)$ & $\Gamma_f(3/2^-)$ & $C_f$ & $\Gamma_f(1/2^+)$ \\
 \hline
 $\Gamma_i(3/2^-)$ &      5.54E+00  &  1.38E-04  &  3.06E-01  & -7.12E-06  \\
$\Gamma_f(3/2^-)$ &      1.38E-04  &  4.39E-09  &  9.02E-06  & -1.34E-10  \\
 $C_f$ &        3.06E-01  &  9.02E-06  &  1.98E-02  & -2.66E-07 \\
$\Gamma_f(1/2^+)$&    -7.12E-06  & -1.34E-10  & -2.66E-07   & 1.76E-11  \\
&&&&&&&&& \\
  $\tag$ &$\Gamma_i(1/2^+)$ & $\Gamma_f(1/2^+ \rightarrow 3/2^-)$ & $C_f$ & $\Gamma_f(1/2^+ \rightarrow 1/2^-)$ \\
\end{tabular}
\end{table}

\newpage

\begin{table}[h]
\caption{$S$ factors at zero energy (or $\sige$ for neutron-capture reactions).}
\begin{tabular}{ll}
Reaction & $S(0)$ or $\sige$ \\ \hline
\dpg\ & $0.223 \pm 0.010$ eV-b (E1:$0.089\pm 0.004$, M1:$0.134\pm 0.006$) \\
\ddn\ & $52.4 \pm 3.5$ keV-b \\
\ddp\ & $57.1 \pm 0.8$ keV-b \\
\tdn\ & $11.7 \pm 0.2$ MeV-b \\
\tag\ & $0.095 \pm 0.005$ keV-b \\
\henp\ & $0.70 \pm 0.01$ MeV$^{1/2}$-b \\
\hedp\ & $5.9 \pm 0.3$ MeV-b \\
\heag\ & $0.51 \pm 0.04$ keV-b \\
\lipa\ & $67 \pm 4$ keV-b \\
\benp\ & $5.75 \pm 0.04$ MeV$^{1/2}$-b \\
\end{tabular}
\end{table}

\newpage
\renewcommand{\baselinestretch}{1.0}
\setcounter{table}{0}
\renewcommand{\thetable}{IV.\alph{table}}
\begin{table}[h]
\caption{$S$-factors}
\begin{tabular}{cccccccccc}
  $E$    &        $\dpg$ &   $E$    &        $\ddn$ &   $E$   &        $\ddp$ &   $E$     &        $\tdn$ &   $E$    &       $\tag$   \\
  (MeV)    &        (eV-b) &   (MeV)    &       (MeV-b) &  (MeV)  &       (MeV-b)&  (MeV)    &        (MeV-b) &   (MeV)   &    (keV-b)      \\ \hline
    0.001 &  2.29E-01  &    0.001 &  5.29E-02  &    0.001 &  5.73E-02  &    0.001 &   1.19E01  &    0.001 &  9.47E-02   \\
   0.002 &  2.35E-01  &    0.002 &  5.33E-02  &    0.002 &  5.75E-02  &    0.002 &   1.21E01  &    0.002 &  9.46E-02   \\
   0.005 &  2.50E-01  &    0.005 &  5.43E-02  &    0.005 &  5.79E-02  &    0.005 &   1.26E01  &    0.005 &  9.43E-02   \\
   0.010 &  2.76E-01  &    0.010 &  5.60E-02  &    0.010 &  5.86E-02  &    0.010 &   1.38E01  &    0.010 &  9.38E-02   \\
   0.020 &  3.29E-01  &    0.020 &  5.95E-02  &    0.020 &  6.01E-02  &    0.020 &   1.71E01  &    0.020 &  9.28E-02   \\
   0.050 &  4.98E-01  &    0.050 &  7.05E-02  &    0.050 &  6.48E-02  &    0.050 &   2.69E01  &    0.050 &  8.98E-02   \\
\end{tabular}
\end{table}

\newpage
\renewcommand{\baselinestretch}{1.0}
\begin{table}[h]
\caption{$S$-factors}
\begin{tabular}{cccccccccc}
  $E$     &      $\henp$ &     $E$    &      $\hedp$ &   $E$     &      $\heag$ &   $E$     &      $\lipa$ &    $E$     &      $\benp$  \\
   (MeV)   &    (b-MeV$^{1/2}$) &   (MeV)    &     (MeV-b) &    (MeV)  &     (keV-b) &  (MeV)    &      (MeV-b) &  (MeV)   &      (b-MeV$^{1/2}$) \\\hline
  0.001 &  6.86E-01  &    0.001 &   5.92E00  &    0.001 &  5.10E-01  &    0.001 &  6.66E-02  &    0.001 &   4.77E00   \\
   0.002 &  6.73E-01  &    0.002 &   5.95E00  &    0.002 &  5.10E-01  &    0.002 &  6.66E-02  &    0.002 &   4.44E00   \\
   0.005 &  6.46E-01  &    0.005 &   6.03E00  &    0.005 &  5.09E-01  &    0.005 &  6.69E-02  &    0.005 &   3.87E00   \\
   0.010 &  6.18E-01  &    0.010 &   6.19E00  &    0.010 &  5.07E-01  &    0.010 &  6.79E-02  &    0.010 &   3.35E00   \\
   0.020 &  5.80E-01  &    0.020 &   6.51E00  &    0.020 &  5.04E-01  &    0.020 &  6.91E-02  &    0.020 &   2.79E00   \\
   0.050 &  5.16E-01  &    0.050 &   7.67E00  &    0.050 &  4.94E-01  &    0.050 &  7.20E-02  &    0.050 &   2.03E00   \\
\end{tabular}
\end{table}

\setcounter{table}{0}
\renewcommand{\thetable}{ V}
\begin{table}[h]
\caption{Analytical fits of the adopted reaction rates.}
\begin{tabular}{lllllll}
 Reaction & $C_0$ & $d_0$ &$d_1$ & $d_2$ & $d_3$ & $T_9^{max}$  \\ \hline
$\dpg$ & 3.7208 & 2.173E+03&	6.899&	-4.442&	3.134&	0.8\\
$\ddn$ & 4.2586 & 4.371E+08	&1.737&	-0.633&	0.109	&3\\
$\ddp$ & 4.2586 & 4.682E+08	&0.745&	-0.065&	0.003&	3\\
$\tdn$ & 4.5244 &8.656E+10&	14.002&	-59.683&	64.236&	0.5\\
$\tag$ & 8.0805 & 7.717E+05	&-0.268&	0.068&	-0.004&	8\\
$\henp$ &   & 6.505E+08	&-0.655	&0.445&	-0.082&	3\\
$\hedp$ & 7.1820 & 5.477E+10&	4.367&	-4.329&	1.115	&2\\
$\heag$ & 12.827 & 5.216E+06&	-0.235&	0.041&	-0.002&	8\\
$\lipa$& 8.4727 & 8.309E+08	&0.278&	-0.018&	0.005&	7\\
$\benp$ &   & 4.609E+09	&-7.518	&53.093	&-135.953&	0.2\\
\end{tabular}
\end{table}

\newpage
\setcounter{table}{0}
\renewcommand{\thetable}{VI.\alph{table}}
\renewcommand{\baselinestretch}{0.95}
\begin{table}[h]
\setcounter{table}{\value{table}}
\caption{$\dpg$}
\begin{tabular}{rlllllll}
 $T_9$ & $E_0$ & $\Delta E_0$ &Adopted & Lower & Upper & NACRE & SKM \\ \hline
0.001 &    0.001 &    0.001 & 1.438E-11  & 1.395E-11  & 1.480E-11  &     1.11 &     0.75 \\
0.002 &    0.002 &    0.001 & 1.991E-08  & 1.933E-08  & 2.047E-08  &     1.11 &     0.76 \\
0.003 &    0.002 &    0.002 & 6.452E-07  & 6.271E-07  & 6.628E-07  &     1.11 &     0.76 \\
0.004 &    0.003 &    0.002 & 5.712E-06  & 5.556E-06  & 5.863E-06  &     1.11 &     0.76 \\
0.005 &    0.003 &    0.003 & 2.674E-05  & 2.603E-05  & 2.743E-05  &     1.11 &     0.76 \\
0.006 &    0.004 &    0.003 & 8.642E-05  & 8.418E-05  & 8.859E-05  &     1.11 &     0.76 \\
0.007 &    0.004 &    0.004 & 2.198E-04  & 2.142E-04  & 2.252E-04  &     1.11 &     0.77 \\
\end{tabular}
\end{table}

\begin{table}[h]
\caption{$\ddn$}
\begin{tabular}{rlllllll}
 $T_9$ & $E_0$ & $\Delta E_0$ &Adopted & Lower & Upper & NACRE & SKM \\ \hline
0.001 &    0.001 &    0.001 & 1.334E-08  & 1.247E-08  & 1.420E-08  &     0.96 &     1.04 \\
0.002 &    0.002 &    0.001 & 5.545E-05  & 5.193E-05  & 5.898E-05  &     0.96 &     1.04 \\
0.003 &    0.003 &    0.002 & 3.064E-03  & 2.873E-03  & 3.254E-03  &     0.97 &     1.04 \\
0.004 &    0.003 &    0.002 & 3.793E-02  & 3.562E-02  & 4.024E-02  &     0.97 &     1.05 \\
0.005 &    0.004 &    0.003 & 2.250E-01  & 2.115E-01  & 2.385E-01  &     0.97 &     1.05 \\
0.006 &    0.004 &    0.003 & 8.689E-01  & 8.175E-01  & 9.203E-01  &     0.97 &     1.05 \\
0.007 &    0.005 &    0.004 &  2.544E00  &  2.394E00  &  2.693E00  &     0.97 &     1.05 \\
\end{tabular}
\end{table}

\begin{table}[h]
\caption{$\ddp$}
\begin{tabular}{rlllllll}
 $T_9$ & $E_0$ & $\Delta E_0$ &Adopted & Lower & Upper & NACRE & SKM \\ \hline
0.001 &    0.001 &    0.001 & 1.438E-08  & 1.418E-08  & 1.457E-08  &     1.02 &     1.05 \\
0.002 &    0.002 &    0.001 & 5.954E-05  & 5.874E-05  & 6.034E-05  &     1.02 &     1.06 \\
0.003 &    0.003 &    0.002 & 3.273E-03  & 3.230E-03  & 3.316E-03  &     1.02 &     1.06 \\
0.004 &    0.003 &    0.002 & 4.034E-02  & 3.981E-02  & 4.087E-02  &     1.02 &     1.06 \\
0.005 &    0.004 &    0.003 & 2.386E-01  & 2.355E-01  & 2.417E-01  &     1.02 &     1.06 \\
0.006 &    0.004 &    0.003 & 9.199E-01  & 9.081E-01  & 9.318E-01  &     1.02 &     1.06 \\
0.007 &    0.005 &    0.004 &  2.692E00  &  2.657E00  &  2.726E00  &     1.02 &     1.05 \\
\end{tabular}
\end{table}

\begin{table}[h]
\caption{$\tdn$}
\begin{tabular}{rlllllll}
 $T_9$ & $E_0$ & $\Delta E_0$ &Adopted & Lower & Upper & NACRE & SKM \\ \hline
0.001 &    0.001 &    0.001 & 1.986E-07  & 1.968E-07  & 2.006E-07  &     1.05 &     1.17 \\
0.002 &    0.002 &    0.001 & 1.437E-03  & 1.424E-03  & 1.452E-03  &     1.04 &     1.10 \\
0.003 &    0.003 &    0.002 & 1.041E-01  & 1.031E-01  & 1.051E-01  &     1.03 &     1.07 \\
0.004 &    0.003 &    0.002 &  1.531E00  &  1.517E00  &  1.546E00  &     1.03 &     1.05 \\
0.005 &    0.004 &    0.003 &  1.029E01  &  1.019E01  &  1.039E01  &     1.03 &     1.04 \\
0.006 &    0.004 &    0.003 &  4.380E01  &  4.340E01  &  4.423E01  &     1.02 &     1.03 \\
0.007 &    0.005 &    0.004 &  1.389E02  &  1.376E02  &  1.403E02  &     1.01 &     1.02 \\
\end{tabular}
\end{table}

\begin{table}[h]
\caption{$\tag$}
\begin{tabular}{rlllllll}
 $T_9$ & $E_0$ & $\Delta E_0$ &Adopted & Lower & Upper & NACRE & SKM \\ \hline
0.001 &    0.002 &    0.001 & 6.297E-28  & 5.888E-28  & 6.583E-28  &   &     0.91 \\
0.002 &    0.004 &    0.002 & 6.893E-21  & 6.445E-21  & 7.205E-21  &     0.94 &     0.93 \\
0.003 &    0.005 &    0.003 & 1.747E-17  & 1.633E-17  & 1.826E-17  &     0.94 &     0.93 \\
0.004 &    0.006 &    0.003 & 2.420E-15  & 2.263E-15  & 2.529E-15  &     0.95 &     0.93 \\
0.005 &    0.007 &    0.004 & 8.012E-14  & 7.494E-14  & 8.374E-14  &     0.94 &     0.94 \\
0.006 &    0.008 &    0.005 & 1.151E-12  & 1.076E-12  & 1.203E-12  &     0.94 &     0.94 \\
0.007 &    0.009 &    0.005 & 9.624E-12  & 9.003E-12  & 1.006E-11  &     0.94 &     0.94 \\
\end{tabular}
\end{table}

\begin{table}[h]
\caption{$\henp$}
\begin{tabular}{rllll}
 $T_9$ & Adopted & Lower & Upper &  SKM \\ \hline
0.001 &    6.796E08  &  6.709E08  &  6.891E08  &      0.96 \\
0.002 &    6.757E08  &  6.672E08  &  6.850E08  &      0.96 \\
0.003 &    6.726E08  &  6.642E08  &  6.818E08  &      0.96 \\
0.004 &    6.699E08  &  6.616E08  &  6.790E08  &      0.96 \\
0.005 &    6.675E08  &  6.593E08  &  6.765E08  &      0.96 \\
0.006 &    6.653E08  &  6.572E08  &  6.742E08  &      0.96 \\
0.007 &    6.633E08  &  6.553E08  &  6.722E08  &      0.96 \\
\end{tabular}
\end{table}

\begin{table}[h]
\caption{$\hedp$}
\begin{tabular}{rllllll}
 $T_9$ & $E_0$ & $\Delta E_0$ &Adopted & Lower & Upper & SKM \\ \hline
0.001 &    0.002 &    0.001 & 3.562E-19  & 3.442E-19  & 3.677E-19  &    0.74 \\
0.002 &    0.003 &    0.002 & 6.156E-13  & 5.948E-13  & 6.353E-13  &    0.81 \\
0.003 &    0.004 &    0.002 & 6.375E-10  & 6.160E-10  & 6.579E-10  &    0.84 \\
0.004 &    0.005 &    0.003 & 5.026E-08  & 4.857E-08  & 5.187E-08  &    0.86 \\
0.005 &    0.006 &    0.004 & 1.116E-06  & 1.078E-06  & 1.151E-06  &    0.87 \\
0.006 &    0.007 &    0.004 & 1.181E-05  & 1.142E-05  & 1.219E-05  &    0.89 \\
0.007 &    0.008 &    0.005 & 7.754E-05  & 7.494E-05  & 8.000E-05  &    0.90 \\
\end{tabular}
\end{table}

\begin{table}[h]
\caption{$\heag$}
\begin{tabular}{rlllllll}
 $T_9$ & $E_0$ & $\Delta E_0$ &Adopted & Lower & Upper & NACRE & SKM \\ \hline
0.001 &    0.004 &    0.001 & 1.035E-47  & 9.753E-48  & 1.095E-47  &   &     1.23 \\
0.002 &    0.006 &    0.002 & 2.027E-36  & 1.910E-36  & 2.143E-36  &   &     1.15 \\
0.003 &    0.008 &    0.003 & 6.010E-31  & 5.663E-31  & 6.357E-31  &   &     1.12 \\
0.004 &    0.009 &    0.004 & 1.688E-27  & 1.590E-27  & 1.785E-27  &   &     1.10 \\
0.005 &    0.011 &    0.005 & 4.766E-25  & 4.491E-25  & 5.041E-25  &     0.96 &     1.08 \\
0.006 &    0.012 &    0.006 & 3.516E-23  & 3.313E-23  & 3.719E-23  &     0.96 &     1.07 \\
0.007 &    0.014 &    0.007 & 1.088E-21  & 1.025E-21  & 1.151E-21  &     0.96 &     1.07 \\
\end{tabular}
\end{table}

\begin{table}[h]
\caption{$\lipa$}
\begin{tabular}{rlllllll}
 $T_9$ & $E_0$ & $\Delta E_0$ &Adopted & Lower & Upper & NACRE & SKM \\ \hline
0.001 &    0.002 &    0.001 & 1.257E-26  & 1.194E-26  & 1.319E-26  &     1.11 &     1.26 \\
0.002 &    0.004 &    0.002 & 3.103E-19  & 2.947E-19  & 3.254E-19  &     1.11 &     1.25 \\
0.003 &    0.005 &    0.003 & 1.170E-15  & 1.111E-15  & 1.226E-15  &     1.11 &     1.24 \\
0.004 &    0.006 &    0.003 & 2.084E-13  & 1.980E-13  & 2.185E-13  &     1.11 &     1.24 \\
0.005 &    0.007 &    0.004 & 8.261E-12  & 7.851E-12  & 8.660E-12  &     1.11 &     1.23 \\
0.006 &    0.008 &    0.005 & 1.362E-10  & 1.294E-10  & 1.427E-10  &     1.11 &     1.23 \\
0.007 &    0.009 &    0.005 & 1.273E-09  & 1.210E-09  & 1.334E-09  &     1.11 &     1.22 \\
\end{tabular}
\end{table}

\begin{table}[h]
\caption{$\benp$}
\begin{tabular}{rllll}
 $T_9$  &Adopted & Lower & Upper & SKM \\ \hline
0.001 &    4.809E09  &  4.784E09  &  4.836E09   &     1.35 \\
0.002 &    4.688E09  &  4.665E09  &  4.714E09   &     1.33 \\
0.003 &    4.599E09  &  4.577E09  &  4.624E09   &     1.32 \\
0.004 &    4.527E09  &  4.505E09  &  4.551E09   &     1.31 \\
0.005 &    4.464E09  &  4.443E09  &  4.488E09   &     1.30 \\
0.006 &    4.409E09  &  4.388E09  &  4.432E09   &     1.30 \\
0.007 &    4.359E09  &  4.339E09  &  4.382E09   &     1.29 \\
\end{tabular}
\end{table}

\setcounter{figure}{0}
\renewcommand{\thefigure}{I.\alph{figure}}
\begin{figure}[h]
\centerline{\epsfig{file=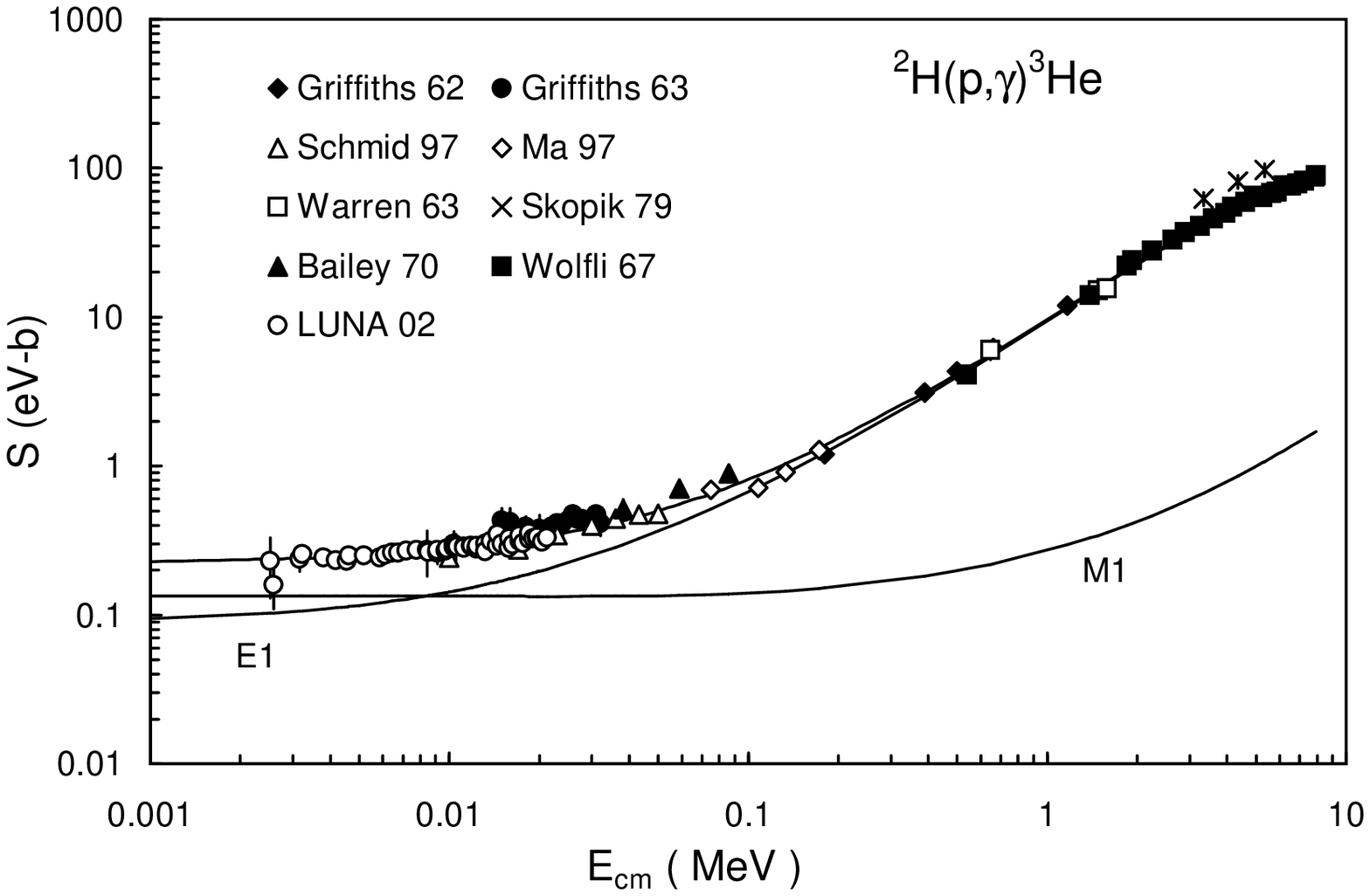,width=14cm,clip=}}
\caption{$\dpg$ $S$ factor. The data are taken from Ref. \protect\cite{Gr62} (Griffiths 62), 
Ref. \protect\cite{Gr63} (Griffiths 63), Ref. \protect\cite{Wa63} (Warren 63), Ref. \protect\cite{WO67} (Wolfli 67),
Ref. \protect\cite{Ba70} (Bailey 70), Ref. \protect\cite{Sk79} (Skopic 79), Ref. \protect\cite{Sc97} (Schmid 97),
Ref. \protect\cite{Ma97} (Ma 97), and Ref. \protect\cite{Lu02} (LUNA 02).
\label{dpg}}
\end{figure}

\begin{figure}[h]
\centerline{\epsfig{file=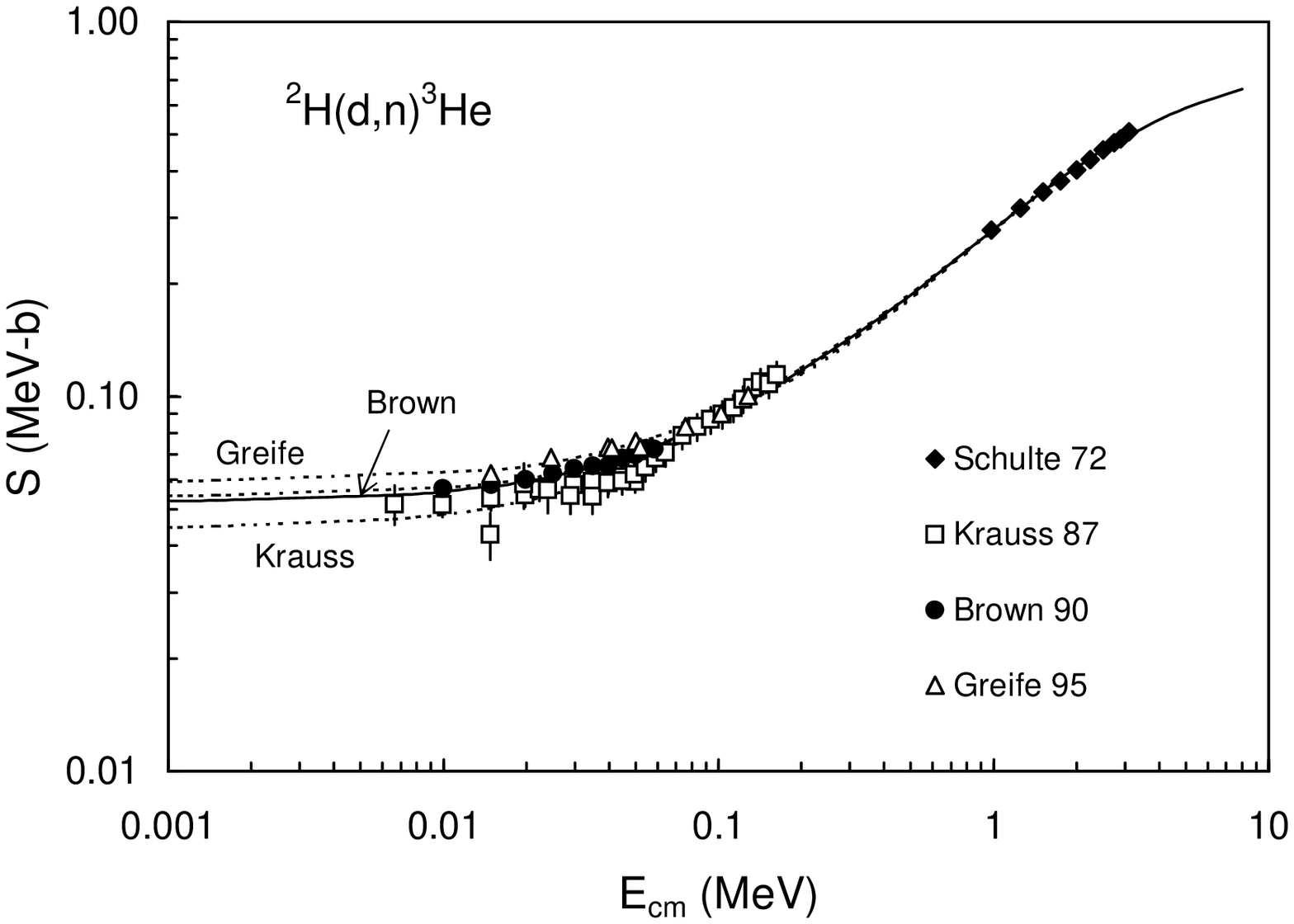,width=14cm,clip=}}
\caption{$\ddn$ $S$ factor. The data are taken from Ref. \protect\cite{SC72} (Schulte 72),
Ref. \protect\cite{KR87a} (Krauss 87), Ref. \protect\cite{BR90} (Brown 90), and Ref. \protect\cite{GR95} (Greife 95). The dotted curves represent the individual fits.
\label{ddn}}
\end{figure}

\begin{figure}[h]
\centerline{\epsfig{file=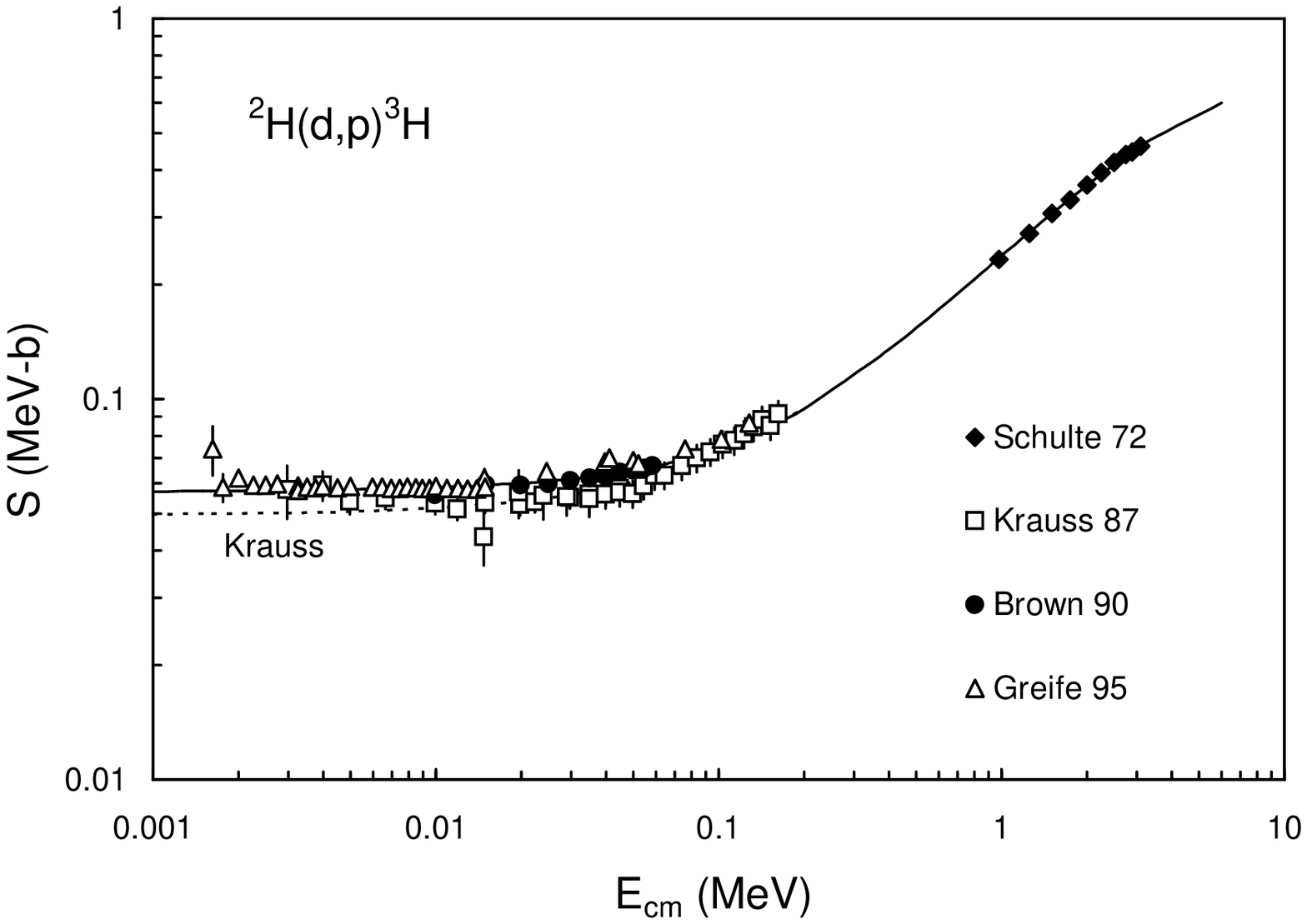,width=14cm,clip=}}
\caption{$\ddp$ $S$ factor. The data are taken from Ref. \protect\cite{SC72} (Schulte 72),
Ref. \protect\cite{KR87a} (Krauss 87), Ref. \protect\cite{BR90} (Brown 90), and Ref. \protect\cite{GR95} (Greife 95). The dotted curve represents the individual fit to Ref.
\protect\cite{KR87a}. Individual fits of Ref. \protect\cite{BR90} and \protect\cite{GR95}
are very close to the global fit.
\label{ddp}}
\end{figure}

\begin{figure}[h]
\centerline{\epsfig{file=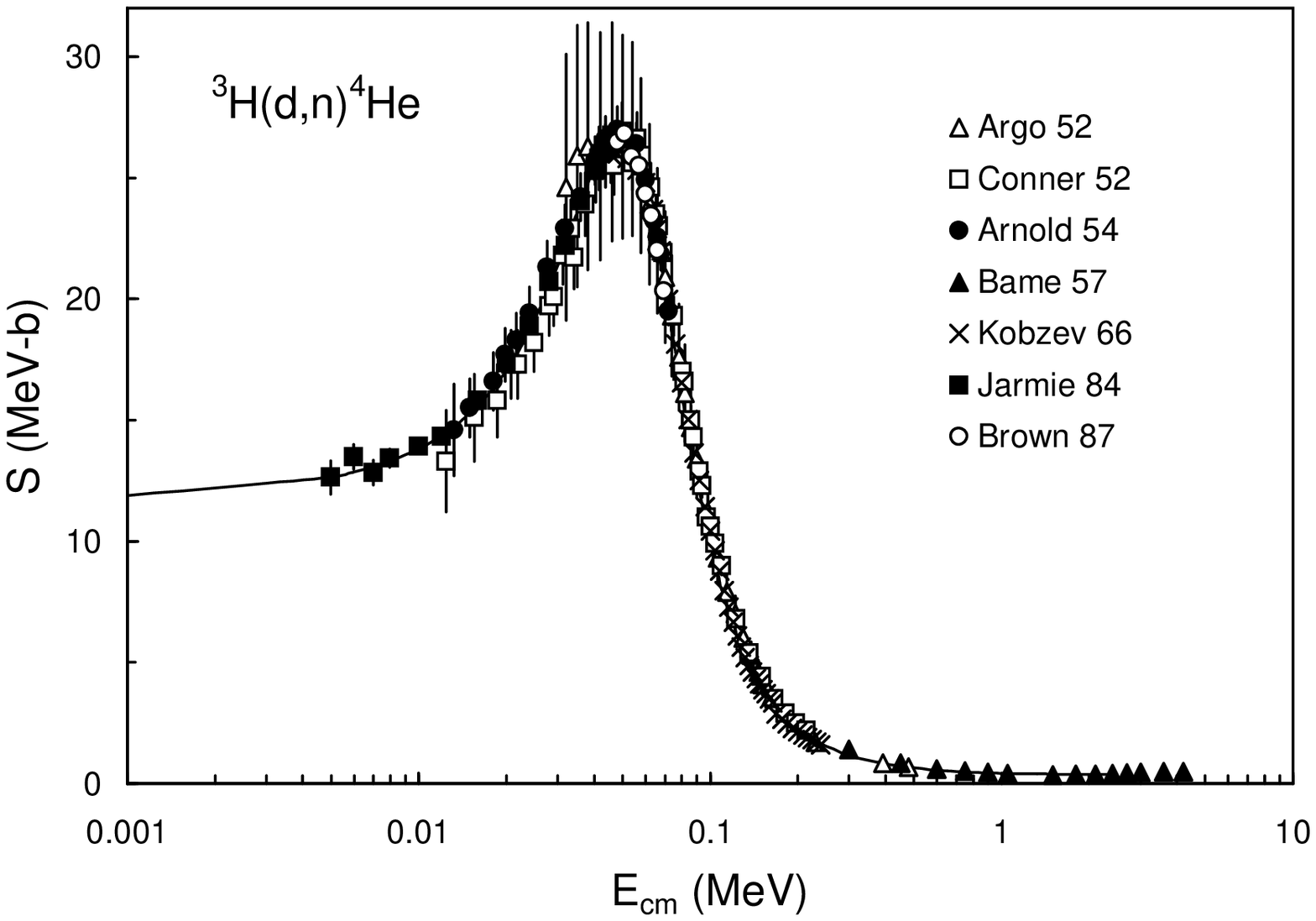,width=14cm,clip=}}
\caption{$\tdn$ $S$ factor. The data are taken from Ref. \protect\cite{AR52} (Argo 52),
Ref. \protect\cite{CO52} (Conner 52), Ref. \protect\cite{AR54} (Arnold 54), Ref. \protect\cite{BA57} (Bame 57), 
Ref. \protect\cite{KO66} (Kobzev 66), Ref. \protect\cite{MC73} (McDaniels), Ref. \protect\cite{JA84} (Jarmie 84),
and Ref. \protect\cite{BR87a} (Brown 87).
\label{tdn}}
\end{figure}

\begin{figure}[h]
\centerline{\epsfig{file=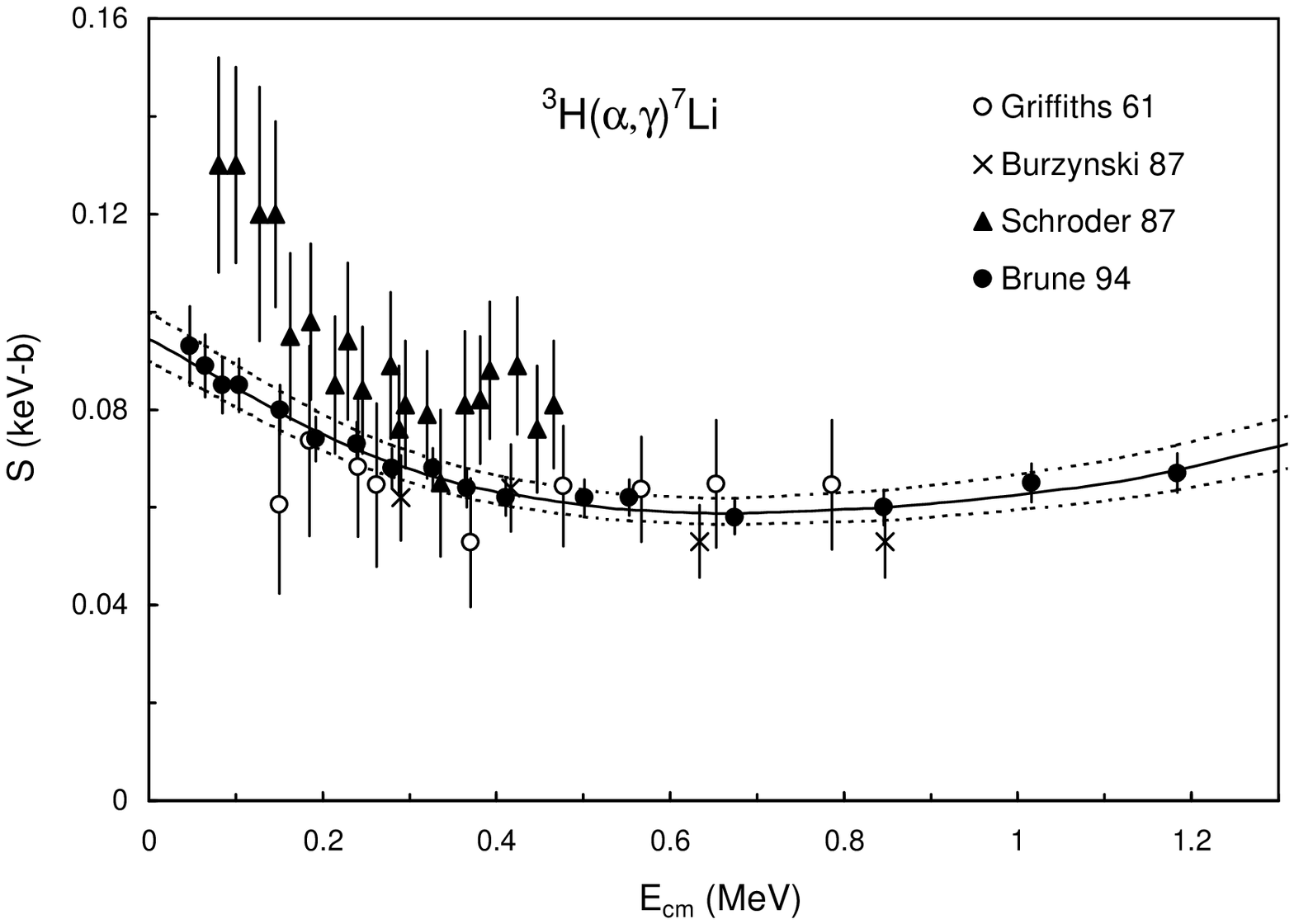,width=14cm,clip=}}
\caption{$\tag$ $S$ factor. The data are taken from Ref. \protect\cite{GR61} (Griffiths 61),
Ref. \protect\cite{BU87} (Burzy\'nski 87), Ref. \protect\cite{SC87a} (Schr\"oder 87), and Ref. \protect\cite{BR94} 
(Brune 94).
\label{tag}}
\end{figure}

\begin{figure}[h]
\centerline{\epsfig{file=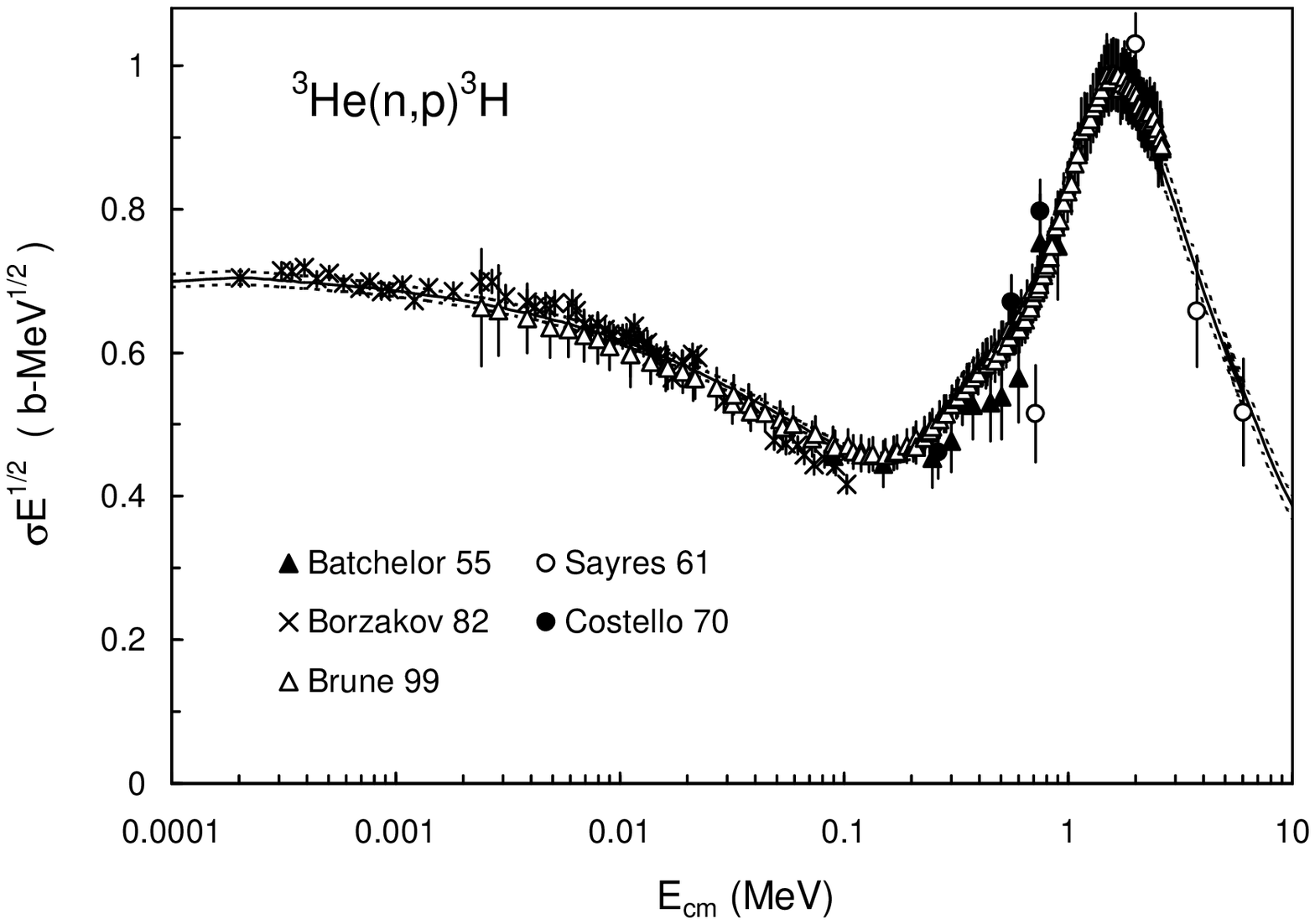,width=14cm,clip=}}
\caption{$\henp$ cross section $(\times \sqrt{E})$.
The data are taken from Ref. \protect\cite{Ba55} (Batchelor 55), 
Ref. \protect\cite{Sa61} (Sayres 61), Ref. \protect\cite{Co70} (Costello 70),
Ref. \protect\cite{Bo82} (Borzakov 82), and Ref. \protect\cite{BHK99} (Brune 99).
\label{he3np}}
\end{figure}

\begin{figure}[h]
\centerline{\epsfig{file=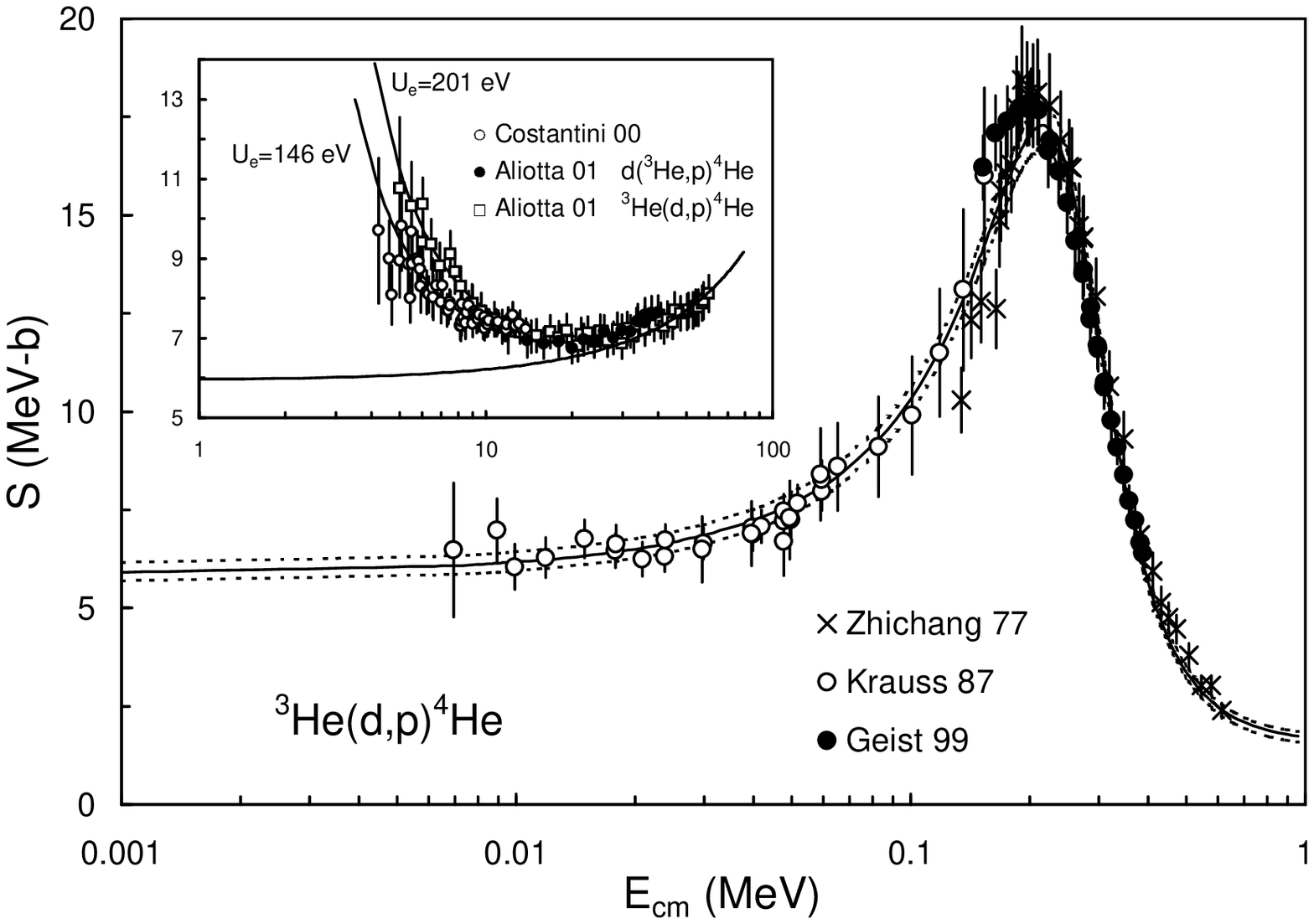,width=14cm,clip=}}
\caption{$\hedp$ $S$ factor. The data are taken from 
Ref. \protect\cite{KR87} (Krauss 87), Ref. \protect\cite{ZH75} (Zhichang 75), 
Ref. \protect\cite{Ge99} (Geist 99), Ref. \protect\cite{Co00} (Costantini 00) and 
Ref. \protect\cite{Al01} (Aliotta 01). The insert shows the influence of electron screening
(energies are given in keV).
\label{he3dp}}
\end{figure}

\begin{figure}[h]
\centerline{\epsfig{file=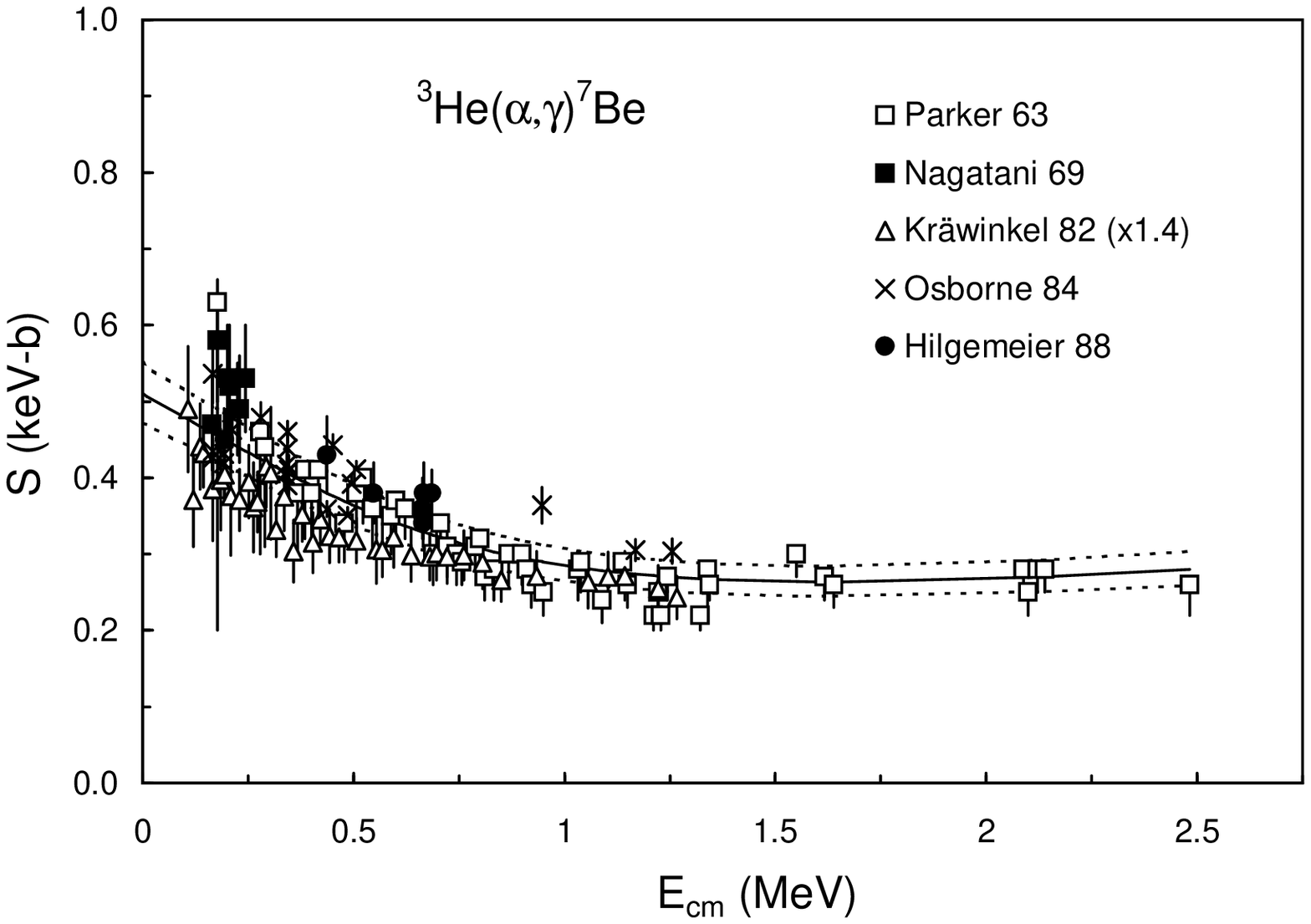,width=14cm,clip=}}
\caption{$\heag$ $S$ factor. The data are taken from Ref. \protect\cite{PA63} (Parker 63),
Ref. \protect\cite{Na69} (Nagatani 69), Ref. \protect\cite{KR82} (Kr\"{a}winkel 82), Ref. \protect\cite{OS84} (Osborne 84),
and Ref. \protect\cite{Hi88} (Hilgemeier 88).
\label{he3ag}}
\end{figure}

\begin{figure}[h]
\centerline{\epsfig{file=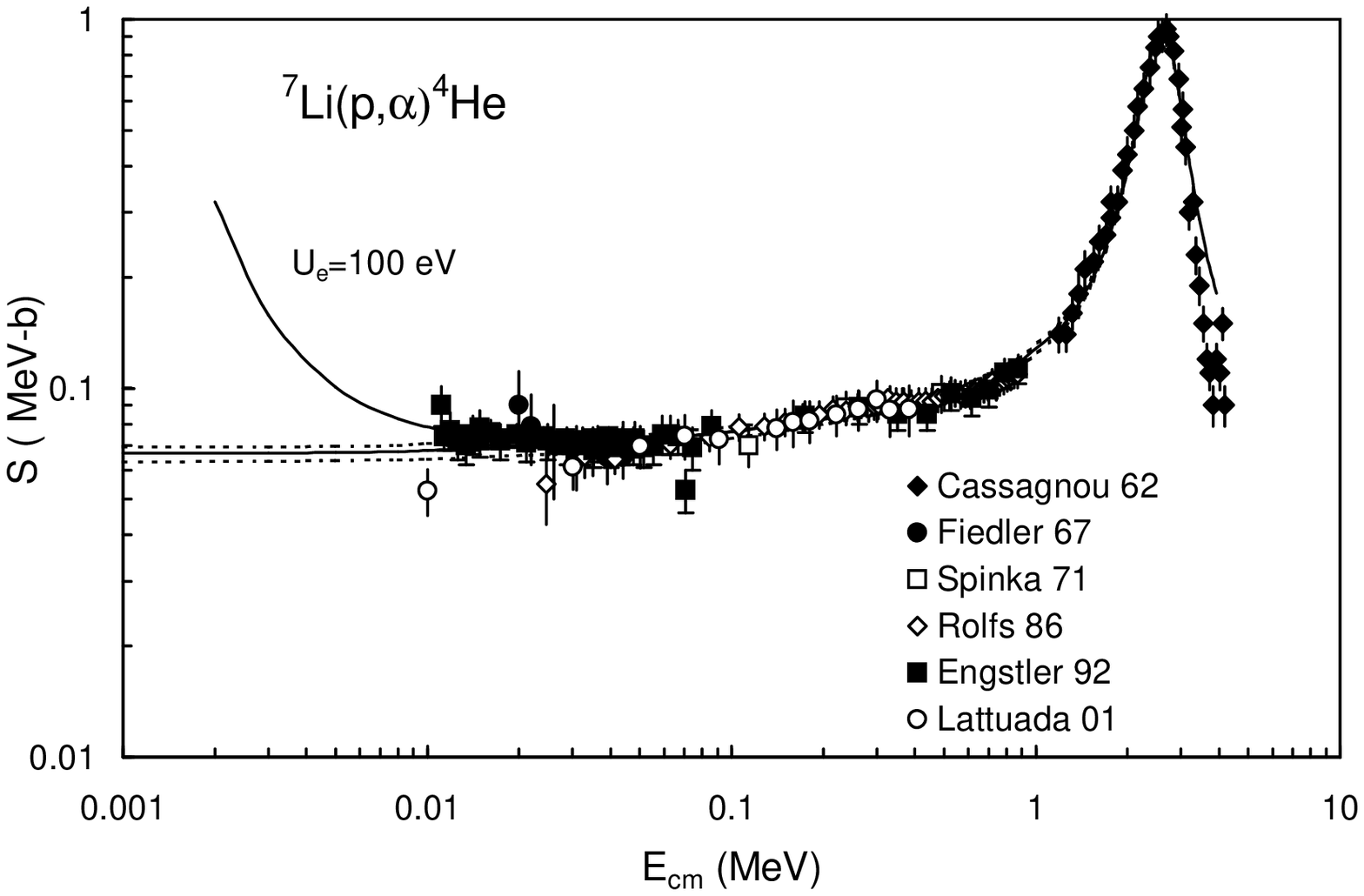,width=14cm,clip=}}
\caption{$\lipa$ $S$ factor. The data are taken from Ref. \protect\cite{CA62} (Cassagnou 62),
Ref. \protect\cite{FI67} (Fiedler 67),
Ref. \protect\cite{SP71} (Spinka 71), Ref. \protect\cite{RO86} (Rolfs 86), Ref. \protect\cite{EN92} (Engstler 92),
and Ref. \protect\cite{LA01} (Lattuada 01).
\label{li7pa}}
\end{figure}

\begin{figure}[h]
\centerline{\epsfig{file=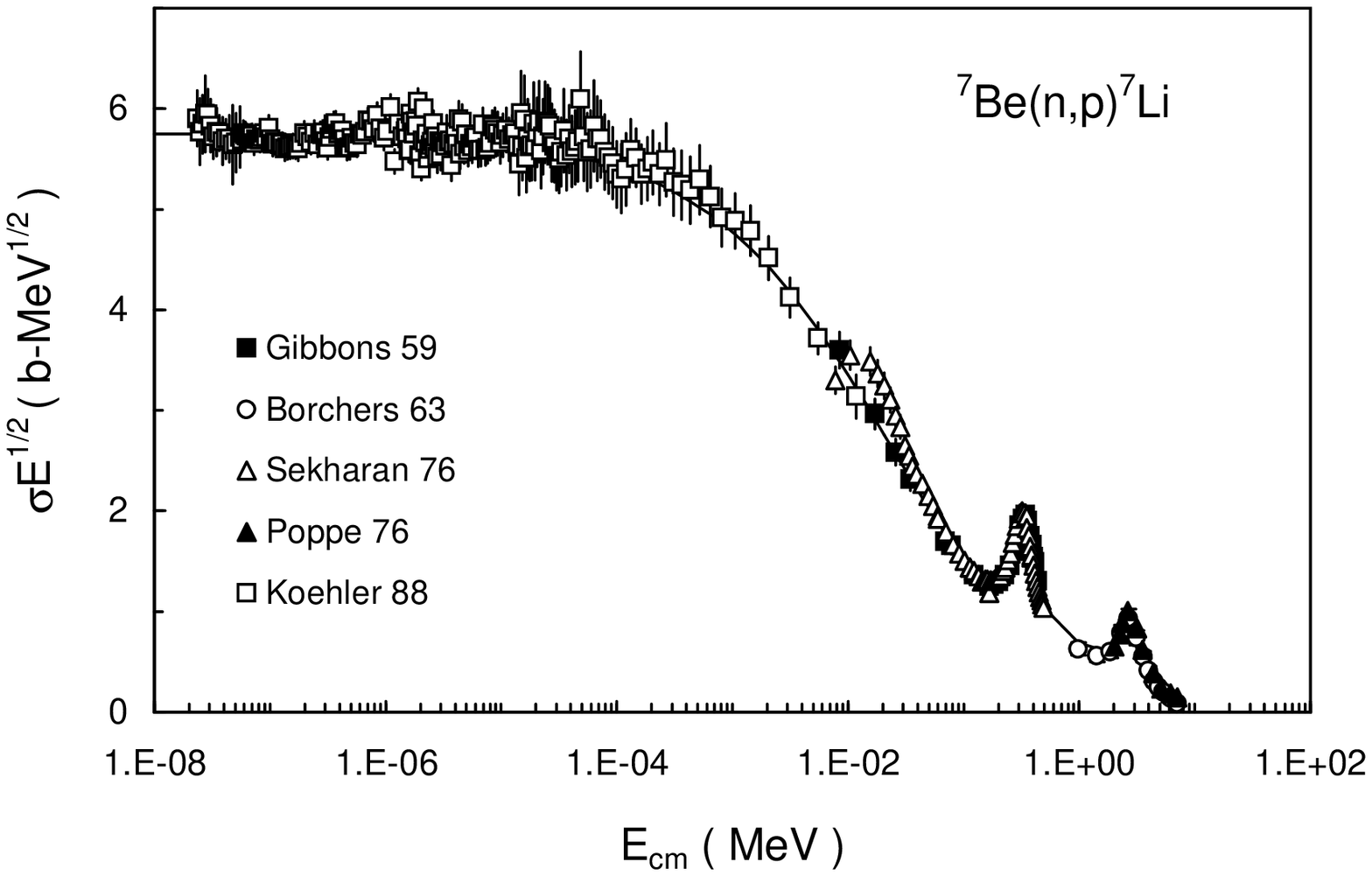,width=14cm,clip=}}
\caption{$\benp$ cross section $(\times \sqrt{E})$.
The data are taken from Ref. \protect\cite{Gi59} (Gibbons 59), Ref. \protect\cite{Bo63} (Borchers 63),
Ref. \protect\cite{Se76} (Sekharan 76), Ref. \protect\cite{Po76} (Poppe 76), and  Ref. \protect\cite{Ko88} (Koehler 88).
\label{be73np}}
\end{figure}

\newpage
\setcounter{figure}{0}
\renewcommand{\thefigure}{II.\alph{figure}}
\begin{figure}[h]
\centerline{\epsfig{file=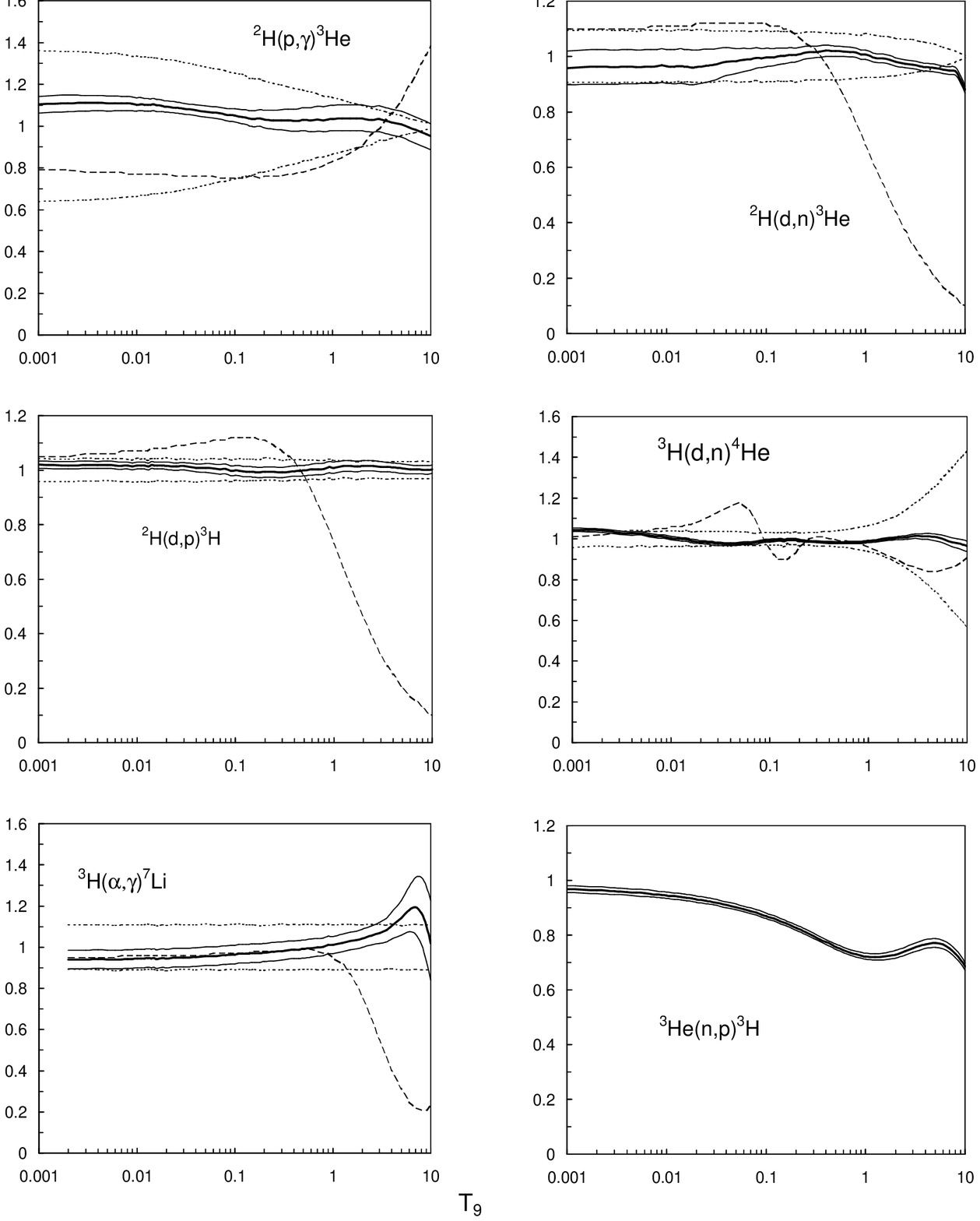,width=18cm,clip=}}
\caption{Ratios of the present reaction rates to the NACRE and SKM rates.
\label{rate1}}
\end{figure}

\begin{figure}[h]
\centerline{\epsfig{file=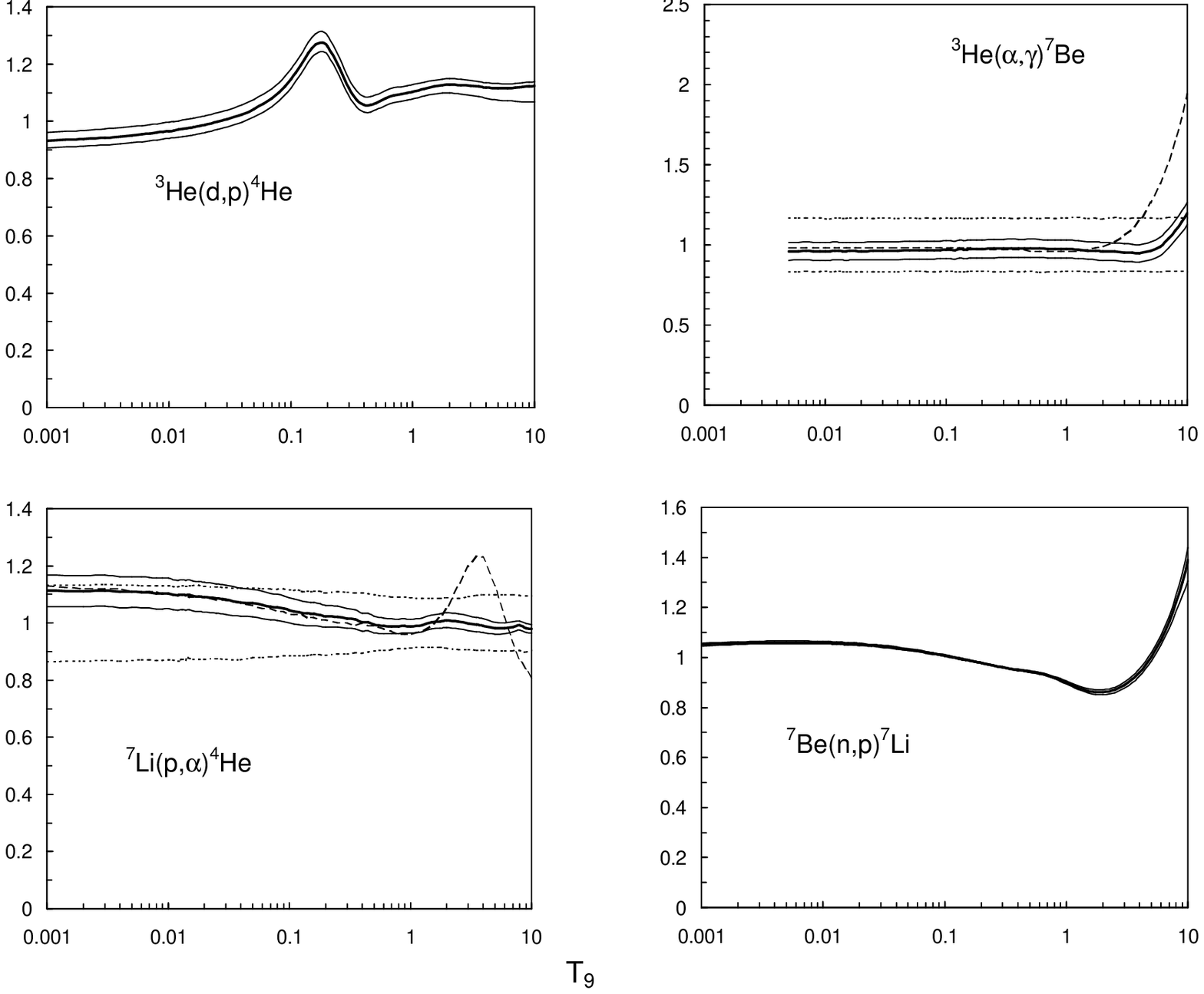,width=18cm,clip=}}
\caption{Ratios of the present reaction rates to the NACRE and SKM rates.
\label{rate2}}
\end{figure}

\end{document}